\documentclass[superscriptaddress,aps,amsmath,amssymb,showpacs,showkeys]{revtex4-2}
\usepackage{natbib} 
\usepackage[dvips]{graphicx}
\usepackage{times}
\usepackage{braket}
\usepackage{xcolor}
\usepackage{orcidlink}
\usepackage{hyperref}
\usepackage{booktabs}
\usepackage{subfigure}
\usepackage{siunitx}
\hypersetup{
	colorlinks=true,
	urlcolor=magenta,
	linkcolor=red,
	citecolor=blue
}

\begin{document}
	\title{Isolated and group environment dependence of stellar mass and different star formation rates}
	
	\author{Pius Privatus\orcidlink{0000-0002-6981-717X}}
	\email[Email: ]{privatuspius08@gmail.com}
	\affiliation{Department of Physics, Dibrugarh University, Dibrugarh 786004, Assam, India}
	\affiliation{Department of Natural Sciences, Mbeya University of Science and Technology, Iyunga 53119, Mbeya, Tanzania}
	
	\author{Umananda Dev Goswami\orcidlink{0000-0003-0012-7549}}
	\email[Email: ]{umananda@dibru.ac.in}
	\affiliation{Department of Physics, Dibrugarh University, Dibrugarh 786004, Assam, India}

	%\date{}
	\begin{abstract}
		In this study, we explored the impact of isolated and group environments on 
		stellar mass, star formation rate (SFR), and specific star formation rate 
		(SSFR) using the catalogue of group and cluster extracted from the Sloan Digital 
		Sky Survey Data Release 12 (SDSS DR12) for $0.02\leq z\leq0.2$. To 
		mitigate the Malmquist bias, we partitioned the entire dataset into eighteen 
		subsamples with a redshift bin size of $\Delta z = 0.01$ and examined the 
		environmental dependencies of these properties within each redshift bin. 
		A strong correlation between environment, stellar mass, SFR, and SSFR was 
		observed across nearly all redshift bins where by the relation is much stronger 
		at lower redshift than higher redshift. 
		The proportional of isolated and group galaxies within the bins was observed 
		to vary with redshift where in the lower redshift bins $(z \lesssim 0.1)$, the 
		proportion of galaxies within the group environment exceeded that within the 
		isolated environment. On the other hand, in the higher redshift bins 
		$(z\gtrsim 0.12)$, the isolated environment's galaxy fraction was found to be 
		higher than that of the group environment. For the intermediate redshift bins 
		$(0.1 \lesssim z \lesssim 0.12)$, an approximately equal proportion of 
		galaxies was observed in both isolated and group environments.  
	\end{abstract}
	
	%\pacs{}
	
	\maketitle    
	\section{Introduction}
	For a long time, it has been established that galaxy formation and evolution 
	are influenced by both internal physics as well as external surroundings. 
	Mechanisms driving internal physics such as halting star formation in 
	galaxies through processes like heating, expulsion, and gas consumption 
	include feedback from active galactic nuclei and supernovae 
	\cite{terrazas2016quiescence}. Beyond internal physics, the external 
	environment in which galaxies exist is anticipated to be another important 
	factor impacting galaxy evolution. Environments with higher density may 
	elevate the merger rates, potentially triggering a rapid quenching process and 
	inducing changes in the shape and star formation rate (SFR) of galaxies on a 
	condensed timescale \cite{peng2010mass}. In most studies, the environment has 
	been characterized by the surface number density $n$ of galaxies within the 
	closest neighbours, falling within the range of $3\leq n \leq 10$ 
	\cite{cooper2006deep2, bag2023shape, yoon2023low, sankhyayan2023identification}.
	As emphasized by \citet{grutzbauch2011galaxy}, the utilization of 
	nearest neighbour densities is widespread in describing the local galaxy 
	environment. The optimal choice for the number of neighbours to consider 
	remains a topic of discussion and relies on the characteristics of the sample 
	and the specific survey \cite{cooper2006deep2,ball2008galaxy,bag2023shape, yoon2023low, sankhyayan2023identification}.
	
	The research conducted by \citet{kauffmann2004environmental} revealed 
	a notable 
	shift of almost a factor of two in the stellar mass distribution 
	of galaxies 
	towards higher masses when comparing low- and high-density regions. 
	Furthermore, 
	the investigation carried out by \citet{etherington2017environmental}
	indicated that 
	high-density environments predominantly have massive galaxies, 
	in contrast 
	to low-density environments. Additionally, \citet{li2006dependence}
	demonstrated 
	that high-mass galaxies exhibit a preference for the densest regions 
	of the Universe, 
	while low-mass galaxies tend to be situated preferentially 
	in low-density 
	regions. In a study focusing on a carefully selected sample of 
	the compact 
	group (CG) of galaxies, \citet{scudder2012dependence}
	noted 
	significant differences in the SFRs of star-forming galaxies
	between the low density and high density. The study by 
	\citet{gomez2003galaxy} demonstrated a robust correlation
	between 
	the SFRs of galaxies and their local (projected) galaxy densities.
	Galaxies 
	situated in dense environments exhibit suppressed SFRs
	\cite{lewis2002df, gomez2003galaxy, tanaka2004environmental, patel2009dependence}.
	According 
	to \citet{lewis2002df}, observing low SFRs extending well beyond 
	the virialised 
	cluster suggests that severe physical processes, such as ram 
	pressure 
	stripping of disk gas, may not be the dominant factor. 
	
	In their investigation, \citet{cooper2008deep2} explored the 
	relationship 
	between the star formation and the environment at both $z \sim 0.1$ and $1$. 
	They reported that in the local Universe,star formation is contingent on the 
	environment, 
	with galaxies in regions of higher over-density generally exhibiting lower SFRs 
	compared
	to their counterparts in lower-density regions. However, \citet{elbaz2007reversal}
	presented contrasting findings in their study, revealing a 
	reversal 
	of the specific star formation-density relation 
	at $z \sim 1$. The average SFR of individual galaxies 
	increased 
	with local galaxy density when the Universe was less than half of 
	its present age. \citet{patel2009dependence} extended this exploration to $z \sim 0.8$ and 
	observed a 
	strong decrease in both the SFR and the specific 
	star formation rate 
	(SSFR, i.e., star formation rate per unit stellar mass) with increasing 
	local density, 
	resembling the relation at $z\sim0$. The study by \citet{xin2021environmental}
	using the AGN galaxies data sample from Sloan Digital Sky Survey, 
	aiming at 
	observing the influence of environment on u-, g-, r-, i- and z-band 
	luminosity, 
	observed that all five luminosities are weakly correlated with galaxy environment further more 
	the study by 
	\citet{deng2023environmental} observed that even u-g, g-r, 
	u-r, i-z and r-i colours of AGN are weakly depending on local environment.  
	
Despite several studies based on galaxy density and field galaxies, 
approaches as mentioned above for the study of variation of galaxy properties 
with the environment are still in debate \cite{rasmussen2012suppression,
ziparo2013lack,wijesinghe2012galaxy,schaefer2017sami,wetzel2013galaxy,
wetzel2014galaxy,mcgee2014overconsumption,peng2015strangulation,
grootes2017galaxy,guglielmo2015star,vulcani2014bluer,oemler2017star,
barsanti2018galaxy,o2023searching}. Using low surface brightness (LSB) 
galaxies \citet{o2023searching} note that there is no clear trend between 
the measured global galaxy properties especially between the galaxy's surface 
brightness and mass. Studying star-forming galaxies in 23 nearby galaxy 
clusters ($z \sim 0.06$) against the field galaxies 
\citet{rasmussen2012suppression} found $\sim 40\%$ decrease in  
SSFR (SSFR $=$ SFR/$M\star$), while \citet{ziparo2013lack} found no decrease in 
SFR and SSFR for the entire population of galaxies in 22 clusters at redshift 
$0 < z < 1.6$. \citet{guglielmo2015star} noted that variations in star 
formation density with the environment are not driven by the differential 
distribution of galaxy mass and morphology in clusters and fields. 
\citet{vulcani2014bluer} shows that galaxy variations in star-forming activity 
and morphology do not depend on the environment. Especially, the most massive 
galaxy of halo or being a satellite, this observation is opposite to 
\citet{oemler2017star,grootes2017galaxy} observations who observed that there 
is an environmental influence. \citet{barsanti2018galaxy} noted a slight 
decrease in the SSFR of star-forming galaxies heading towards the cluster 
center by a factor of $\sim 1.2$ compared to field galaxies. Both 
\citet{schaefer2017sami} and \citet{wijesinghe2012galaxy} examined galaxy 
clusters and field galaxies together but different conclusions have been 
reached. According to \citet{wijesinghe2012galaxy}, the relationship between 
the SFR density and a particular galaxy is known only when the population of 
passive star-forming galaxies is included. This suggests that a galaxy's 
current SFR is largely affected by its stellar mass without any visible impact 
on the environment. On the contrary \citet{schaefer2017sami} found that the 
SFR gradient in star-forming galaxies becomes steeper as total SFR decreases 
in dense environments.
	
	In this study for the first time we use the isolated galaxies, a specific subset of field galaxies with spatial isolation criteria and 
luminosity restrictions, employing the catalogue of group and clusters from SDSS DR 12 by \citet{tempel2017merging} 
	to fully explore the variation of stellar mass, SFR and SSFR based on 
	isolated and group environment approach for the redshift covering $z\le 0.2$. 
	The study 
	will compare the isolated and group environment approach results with the previous results obtained from field galaxy and density approaches.
	
	Our paper is organized as follows: In Section \ref{secII} we present
	the survey 
	from which the data of the galaxy samples are taken. The method of analysis of these data samples is 
	also discussed 
	in this section. Section \ref{secIII} is used to present the results and their 
	discussion. 
	We conclude the study in Section \ref{secIV}. Throughout this paper, we 
	consider the 
	cosmological parameters provided in \citet{collaborartion2016planck}: the Hubble constant 
	$H_0=67.8$ km s$^{-1}$ Mpc$^{-1}$, the matter density $\Omega_{m}=0.308$, 
	and the dark 
	energy density $\Omega_{\Lambda}=0.692$.
	
	\section{Data and Analysis} 
	\label{secII}
	As mentioned in the previous section, this study relies on catalogue data 
	extracted 
	from Sloan Digital Sky Survey Data Release 12 (SDSS DR12) \cite{eisenstein2011sdss, alam2015eleventh}.
	It's important 
	to note that the galaxy data in SDSS DR16 and DR17, which constitute the final data releases of 
	the SDSS IV, 
	remains virtually unchanged compared to DR12 within the Survey area \cite{accetta2022seventeenth}. 
	Specifically, 
	we utilize the value-added catalogue of galaxies, groups, and clusters 
	provided 
	by \citet{tempel2017merging}. The galaxy sample utilized in \citet{tempel2017merging} is 
	volume-limited and 
	contains $584,449$ galaxies with spectroscopic redshifts up to $z = 0.2$, all of which are 
	brighter 
	than the Petrosian r-band magnitude of $17.77$. Additionally,the catalogue includes $88,662$ 
	galaxy groups, each 
	consisting of a minimum of two members. These groups were compiled using a 
	modified friend-of-friend (FoF) method with a variable linking length. The essence of 
	this approach
	lies in the division of the sample into distinct systems through an objective and automated process. 
	The method 
	involves creating spheres with a linking length ($R$) around each sample point, specifically galaxies 
	in this context
	To adjust the linking length based on distance, the procedures outlined in \citet{tempel2014flux,tempel2017merging}
	was applied. 
	The relationship between the linking length and the redshift is represented by an arctangent law as given by
	\begin{equation}
		R_{LL}(z) = R_{LL,0} \left[1 + a \arctan\left(\frac{z}{z_\star}\right)\right],
		\label{eq1} 
	\end{equation}
	where $R_{LL}(z)$ is the linking length used to create a sphere at a specific redshift, $R_{LL,0}$ is 
	the linking length at $z = 0$, $a$
	and ${z_\star}$ are free parameters. The values of $R_{LL,0}=0.34$ Mpc,$a=1.4$ and ${z_\star}=0.09$ are obtained by 
	fitting Equation (\ref{eq1}) 
	to the linking length scaling relation. If there are other galaxies within the sphere, they are considered as the 
	parts of the same system and 
	referred to as ``friends''.  Subsequently, additional spheres are drawn around these newly identified neighbours, 
	and the process 
	continues with
	the principle that ``any friend of my friend is my friend''. This iterative procedure persists until no new 
	neighbours or
	``friends'' can be added. At that point, the process concludes, and a system is defined. Two galaxy groups are 
	said to be 
	interacting if the distance between their centers (in comoving coordinates) is less than the sum of their radii. 
	Consequently, 
	each system comprises either a solitary, isolated galaxy or a group of galaxies that 
	share at
	least one neighbour within a distance not exceeding $R$. For comprehensive details regarding the 
	compilation of the galaxy samples, refer to \citet{tempel2017merging}. 
	In this study, we used the luminous volume-limited main galaxy sample, 
	with r-band absolute magnitude in the range of $ -22.5\leq M_r \leq -20.5$ in which galaxies with above log(M$\star$)$\sim$10 correspond to 
Milk Way-like galaxies (MW). The apparent magnitude was transformed into 
absolute magnitude using the relation:
	\begin{equation}
		M_{r}= m_r-25-5log_{10} (d_L)-K,
		\label{eqR} 
	\end{equation}
	where $d_L$ is the luminosity distance, $M_{r}$ and $m_{r}$ are r-band 
	absolute and apparent magnitudes respectively, and K is the k+e-correction. 
	The k-corrections were calculated with the KCORRECT ($v4\_2$) algorithm 
	\cite{blanton2007k}. The evolution corrections were estimated using the 
	luminosity evolution model of $K_e = c*z$, where $z = -1.62$ for the 
	r-filter \cite{blanton2003galaxy}. The magnitudes correspond to the 
	rest-frame (at the redshift z = 0) and evolution correction was estimated 
	similarly by assuming a distance-independent luminosity function 
	\cite{tempel2012groups,tempel2014flux}.
	
	The data on stellar mass, SFR, and SSFR were derived from
	the 
	spectroscopic catalogues of the SDSS. The analysis 
	utilized 
	the spectra processed by the Max Planck Institute 
	for 
	Astrophysics and Johns Hopkins University (MPA-JHU). 
	The MPA-JHU 
	employs sophisticated stellar population synthesis models to precisely fit and 
	remove 
	the stellar continuum, yielding a collection of
	emission-line 
	measurements for the galaxy spectra. This
	methodology 
	has been applied to earlier releases of SDSS data, 
	and the 
	obtained measurements have been utilized in diverse
	scientific investigations 
	\cite{brinchmann2004physical, kauffmann2004environmental,  tremonti2004origin}.
	The set 
	of line measurements is commonly referred to as the 
	MPA-JHU measurements, 
	named after the Max Planck Institute for 
	Astrophysics 
	and Johns Hopkins University, where the technique
	was developed. 
	These measurements were provided for all objects 
	identified as 
	galaxies by SDSS software, idlspec2d \cite{york2000sloan}.
	This dataset 
	was chosen for its photometric completeness, 
	uniform 
	spectral calibration, broad redshift coverage ($0.02 \leq z \leq 0.6$), 
	and a 
	diverse range of emission line properties associated with
	galaxy 
	formation and evolution, including SFRs, stellar 
	masses (M$\star$), 
	and SSFRs, following the methodologies 
	outlined in \citet{brinchmann2004physical, kauffmann2004environmental,  tremonti2004origin}.
	
	The determination of stellar masses in the MPA-JHU relies on the Bayesian approach 
	and model 
	grids provided by \citet{kauffmann2004environmental}.
	Spectra 
	and photometry are used in conjunction, with a minor
	correction 
	made for the contribution of nebular emission. 
	For the 
	SDSS spectroscopic fiber 
	aperture, 
	stellar mass is computed using fiber magnitudes,
	while the 
	overall stellar mass is calculated using model magnitudes.
	Regarding 
	SFR, as detailed in \citet{brinchmann2004physical}, 
	they are 
	computed within the galaxy fiber aperture using nebular emission lines.
	For regions 
	outside the fiber, SFRs are estimated using galaxy photometry
	data obtained 
	from \citet{salim2015mass}. In cases of AGNs and galaxies
	with weak 
	emission lines, SFRs are estimated solely from photometry. 
	The MPA-JHU data lists the values of stellar mass, SFR and SSFR at the median and 2.5th, 16th, 84th, 97.5th percentiles of the probability distribution function. In this study, the median estimates of the probability distribution function are used to 
	be consistent with the majority of studies \cite{xin2021environmental}.
	
\begin{figure}[!p]
		\centering
		\includegraphics[scale=0.15]{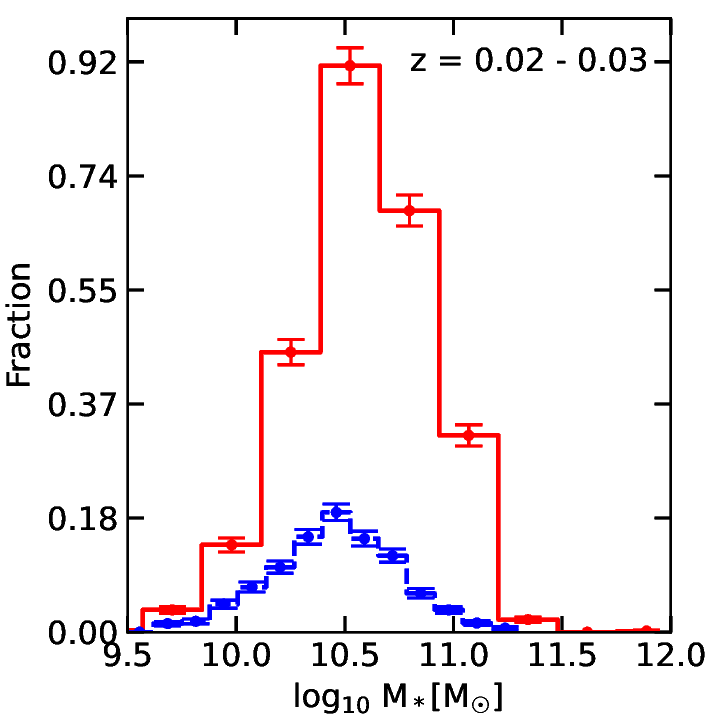}\hspace{0.5cm}
		\includegraphics[scale=0.15]{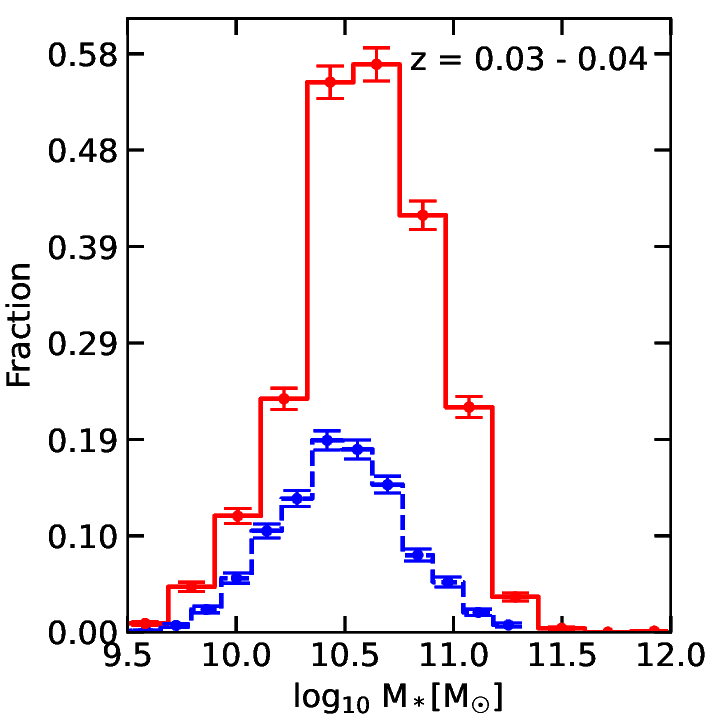}\hspace{0.5cm}
		\vspace{0.5cm}
		\includegraphics[scale=0.15]{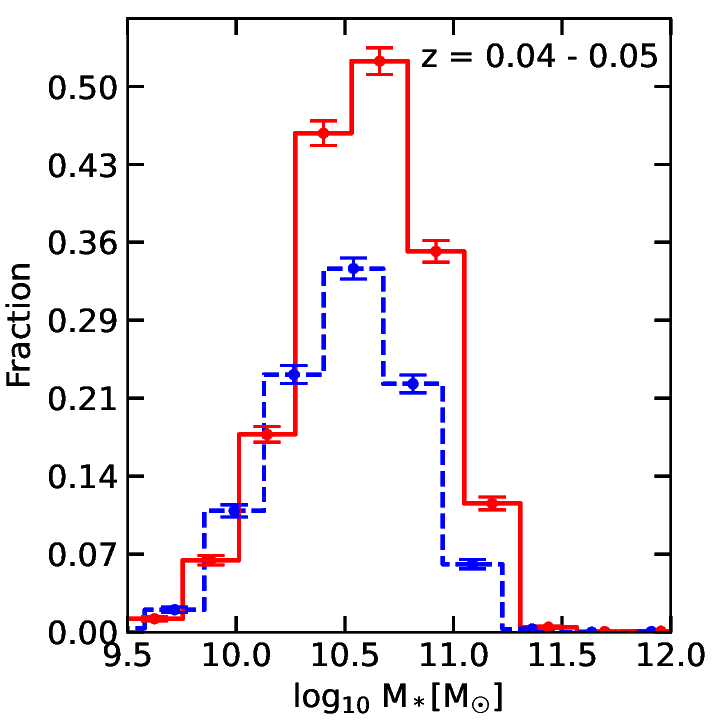}\hspace{0.5cm}
		\includegraphics[scale=0.15]{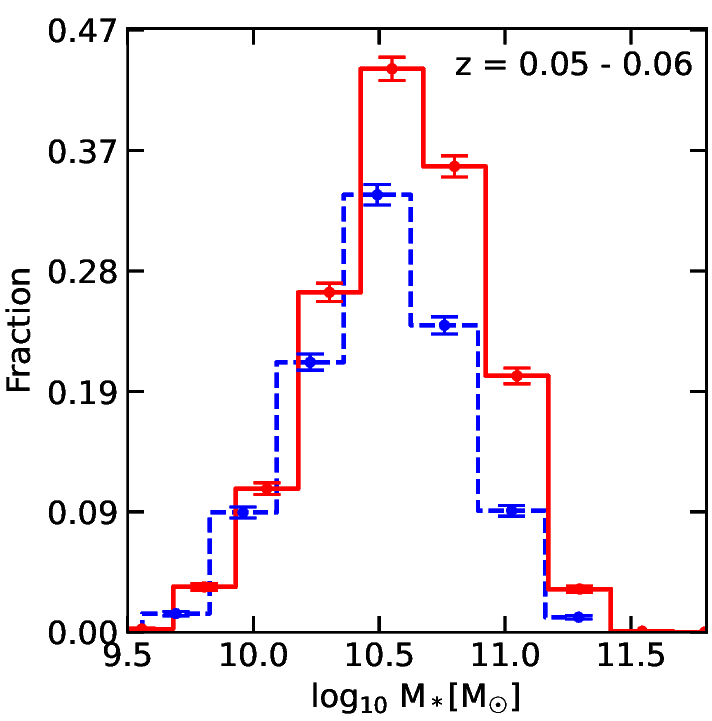}\hspace{0.5cm}
		\includegraphics[scale=0.15]{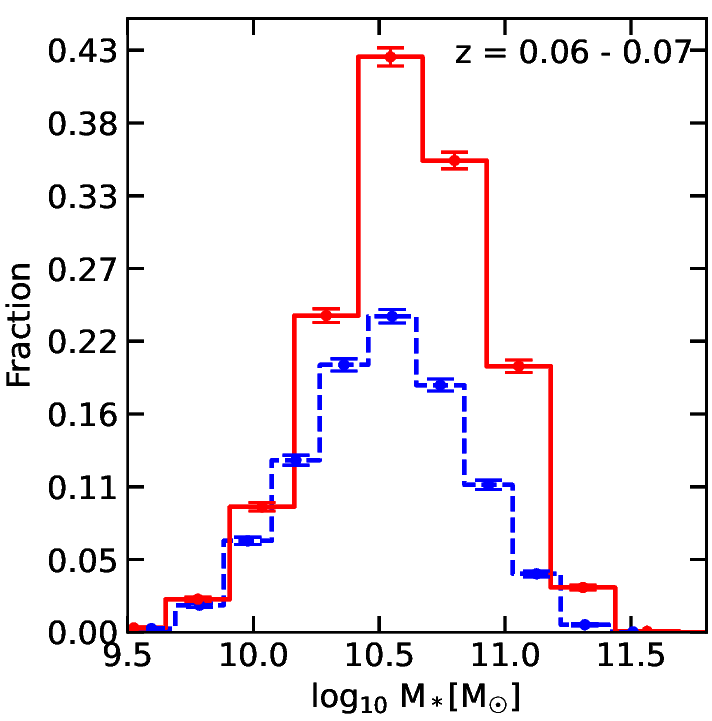}\hspace{0.5cm}
		\includegraphics[scale=0.15]{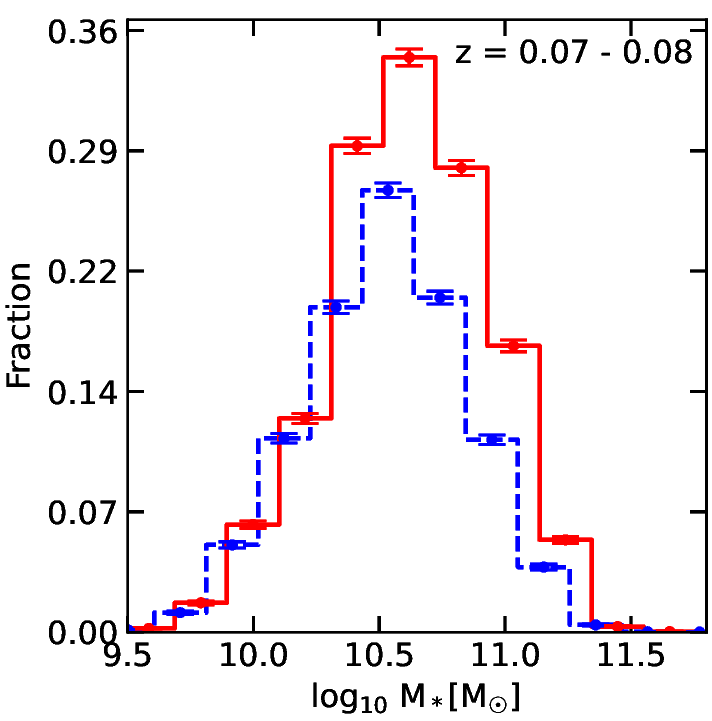}\hspace{0.5cm}
		\vspace{0.5cm}
		\includegraphics[scale=0.15]{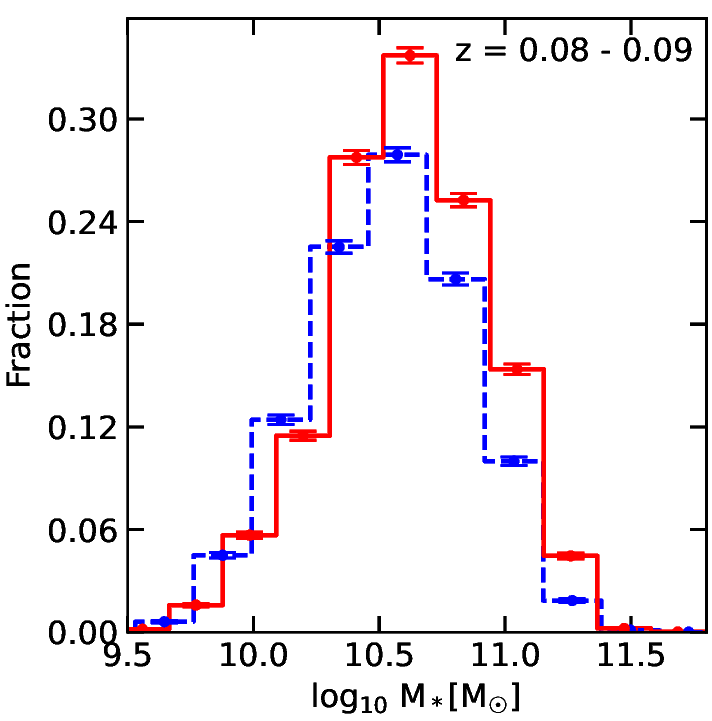}\hspace{0.5cm}
		\includegraphics[scale=0.15]{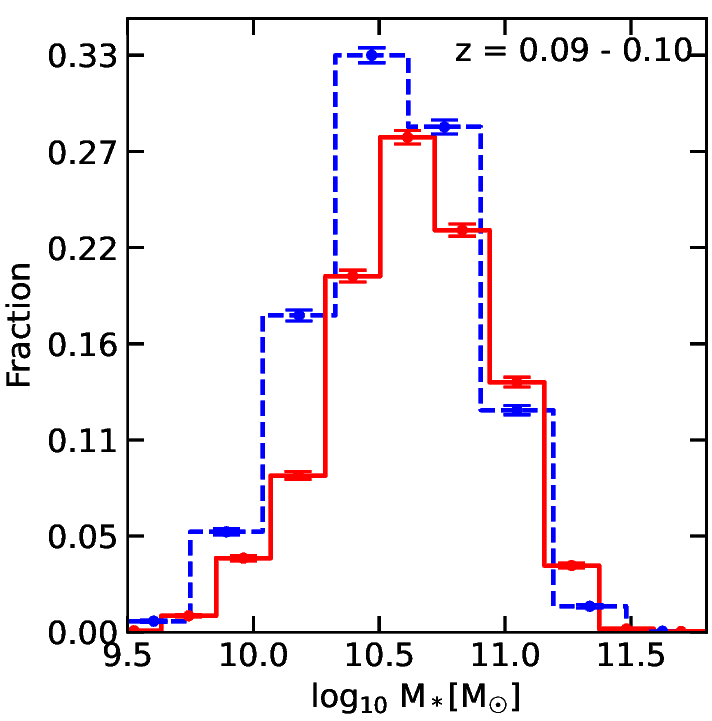}\hspace{0.5cm}
		\includegraphics[scale=0.15]{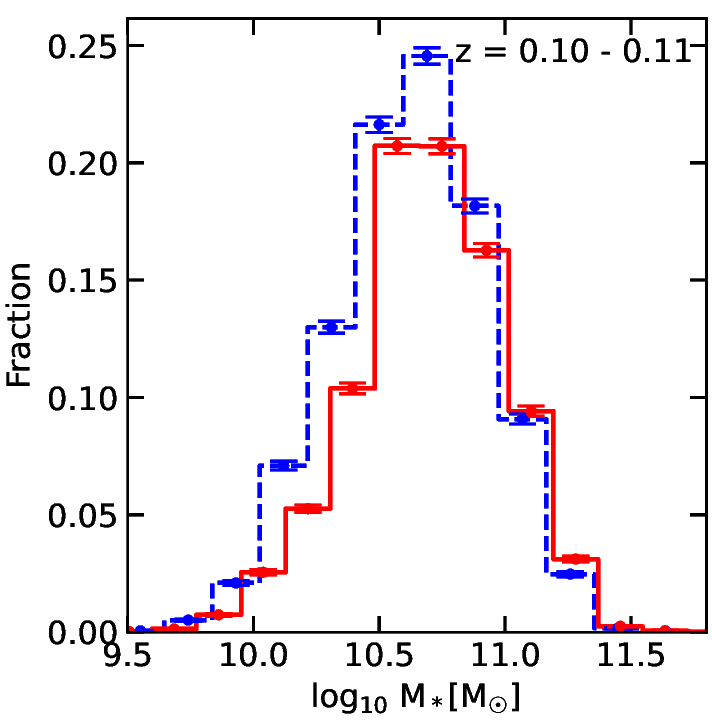}\hspace{0.5cm}
		\includegraphics[scale=0.15]{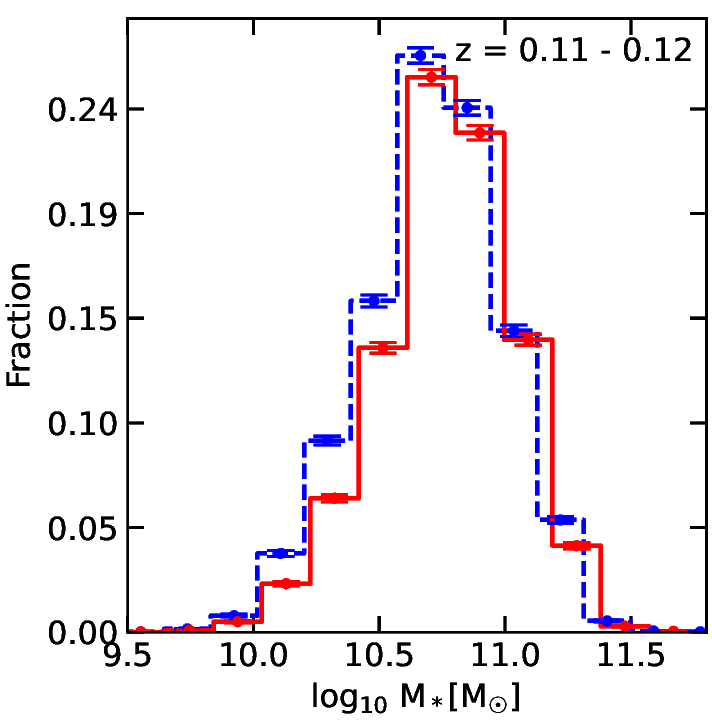}\hspace{0.5cm}
		\vspace{0.5cm}
		\includegraphics[scale=0.15]{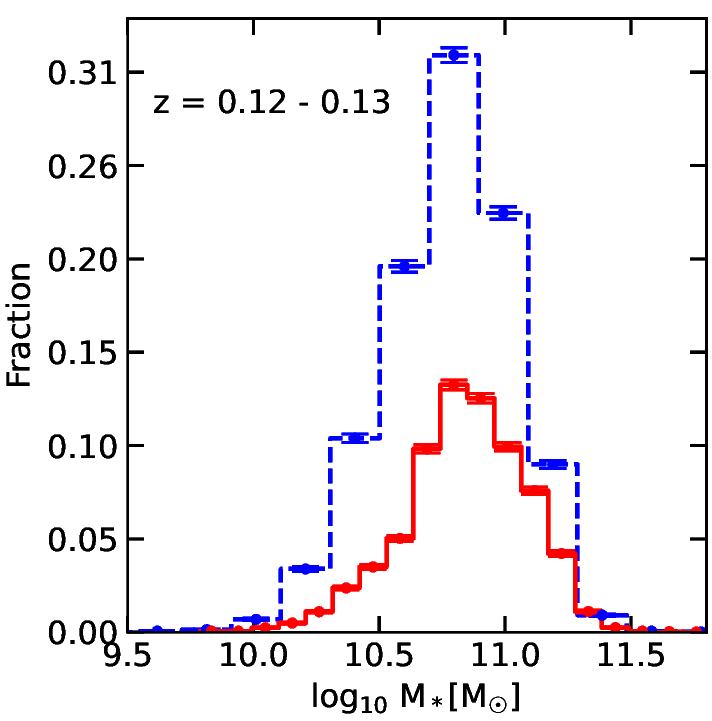}\hspace{0.5cm}
		\includegraphics[scale=0.15]{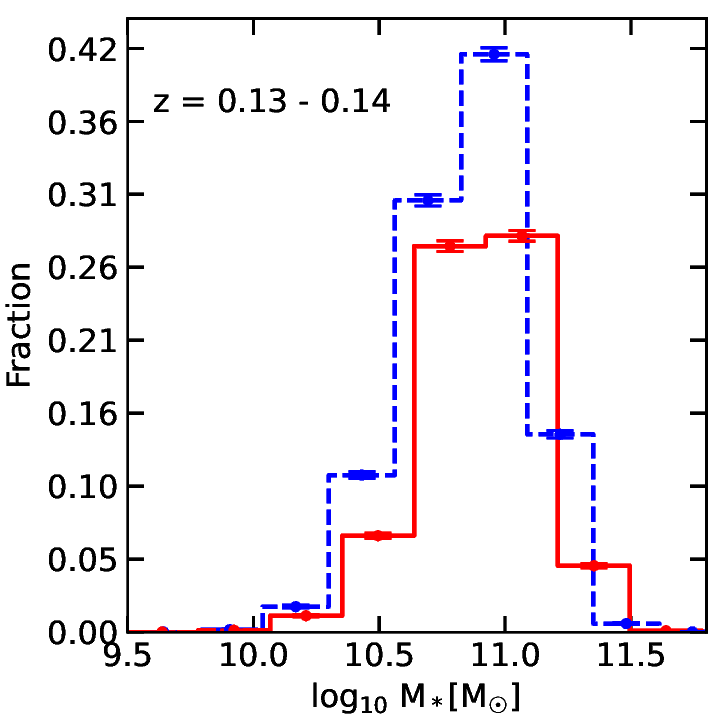}\hspace{0.5cm}
		\includegraphics[scale=0.15]{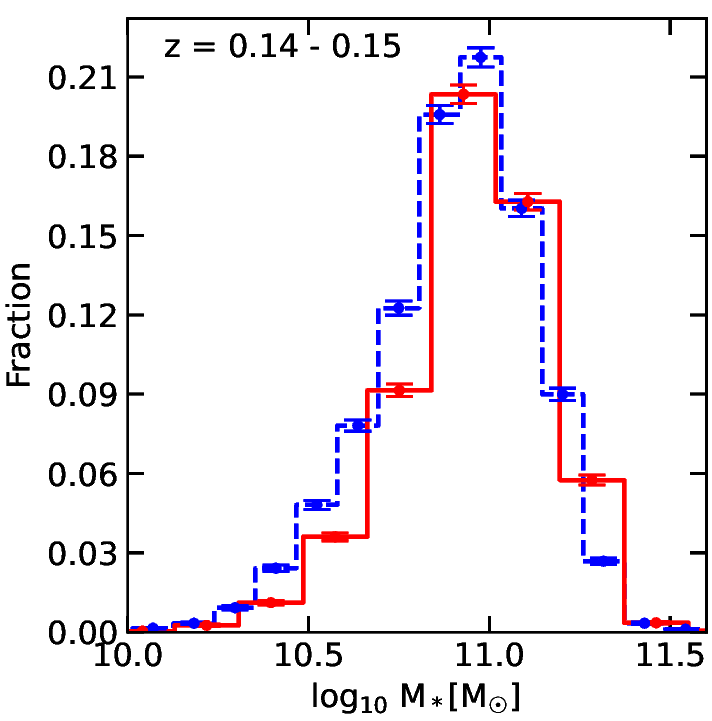}\hspace{0.5cm}
		\includegraphics[scale=0.15]{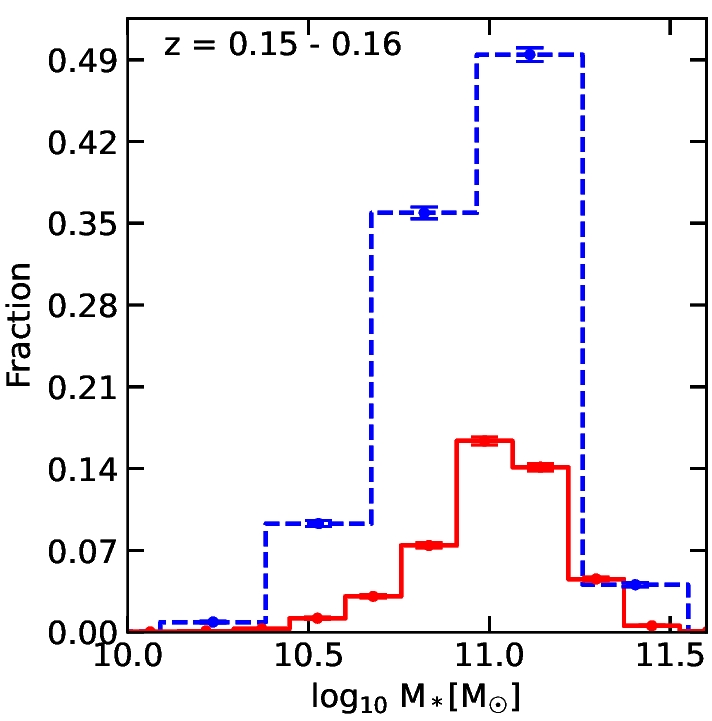}\hspace{0.5cm}
		\vspace{0.5cm}
		\includegraphics[scale=0.15]{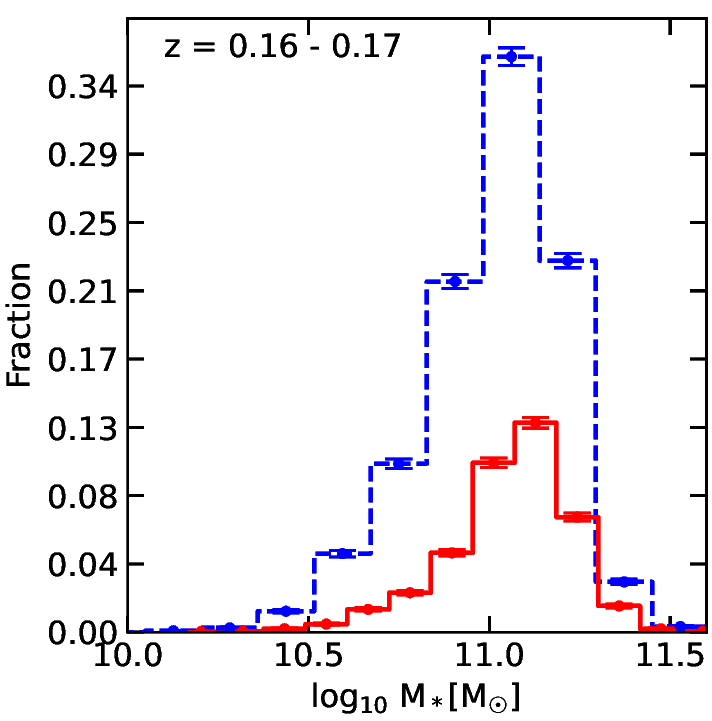}\hspace{0.5cm}
		\includegraphics[scale=0.15]{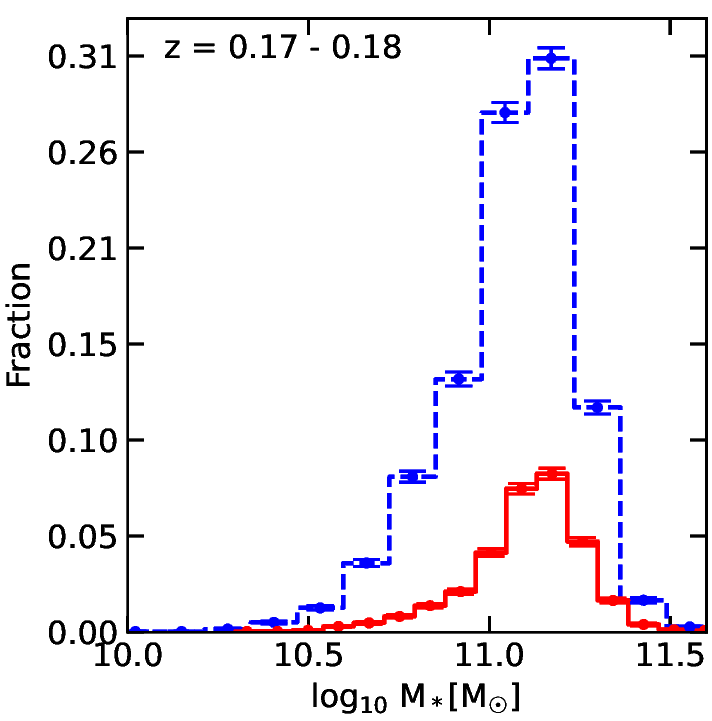}\hspace{0.5cm}
		\includegraphics[scale=0.15]{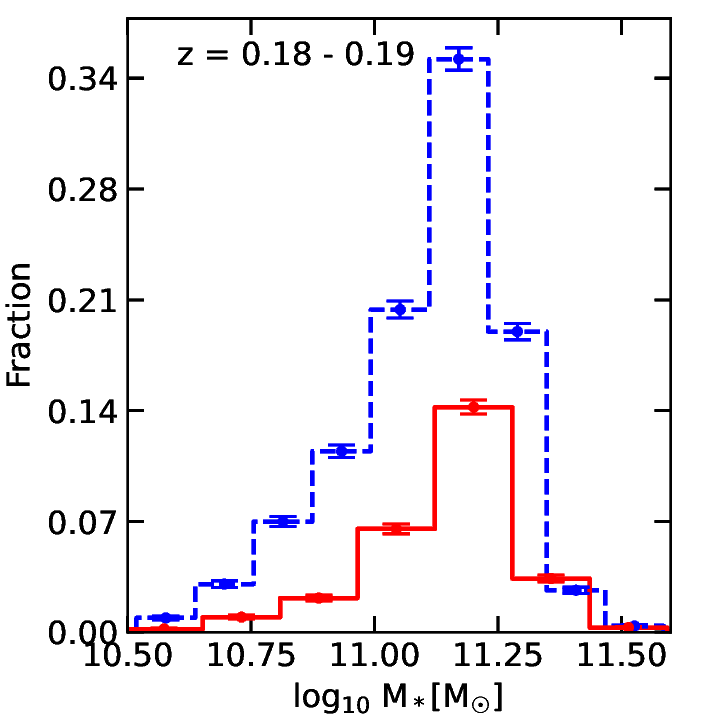}\hspace{0.5cm}
		\includegraphics[scale=0.15]{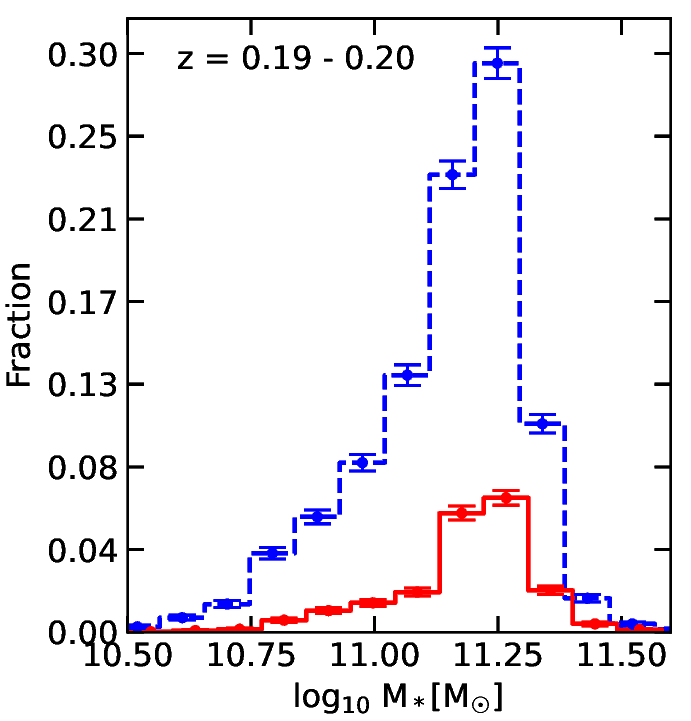}\hspace{0.5cm}
		\caption{Stellar mass distributions for isolated galaxies (blue dotted 
			line) and group galaxies (red solid line). The error bars are 1 $\sigma$ 
			Poissonian errors.}
		\label{stellarmass}
	\end{figure}

	The final dataset comprises a total of $397,408$ galaxies, 
	obtained 
	through a $1$ arcsec sky cross-correlation between the 
	value-added 
	catalogue data of SDSS DR12 by \citet{tempel2017merging}
	and MPA-JHU. 
	The galaxies with \texttt{Group} \texttt{Id} $=0$ are termed 
	isolated, 
	while the galaxies with \texttt{Group} \texttt{Id} $\neq0$ are 
	termed groups. 
	Isolated galaxies represent a population of galaxies with 
	minimized 
	environmental evolutionary effects while Group galaxies represent 
	a population 
	of galaxies with high environmental evolutionary effects. 
	The galaxies 
	were selected within the redshift range of $0.02\leq z \leq 0.2$, making the 
	total 
	number of $397,408\, (100\%)$ where isolated galaxies are $213,535\, (53\%)$ 
	and group 
	galaxies are $183,873\, (46\%)$. It's also important to address the potential 
	bias 
	introduced by the Malmquist effect, wherein there 
	is a 
	tendency for observers to perceive an increase in averaged 
	luminosity 
	with distance due to the non-detection of less luminous
	objects at 
	larger distances \cite{teerikorpi2015eddington}. This
	bias can 
	significantly impact statistical results. To mitigate the Malmquist effect,
	the entire 
	main galaxy sample was divided into subsamples with a 
	redshift 
	binning size of $\Delta z = 0.01$. The minimum cutoff redshift ($z\leq 0.02$) 
	aim to minimize 
	the local effect and using bins with substantial number of galaxies (at least 1000 per bin).  
	This approach 
	ensures that the radial selection function is approximately uniform 
	within each 
	bin, minimizing variations in the space density of galaxies with radial 
	distance 
	and decreasing the impact of the Malmquist bias. Figures \ref{stellarmass}, 
	\ref{sfr} and 
	\ref{ssfr} show the distribution of stellar mass, SFR and SSFR for all redshift bins, respectively.
	
	\section{Results and Discussion}
	\label{secIII}
\begin{figure}[!h]
		\centering{
			\includegraphics[scale=0.21]{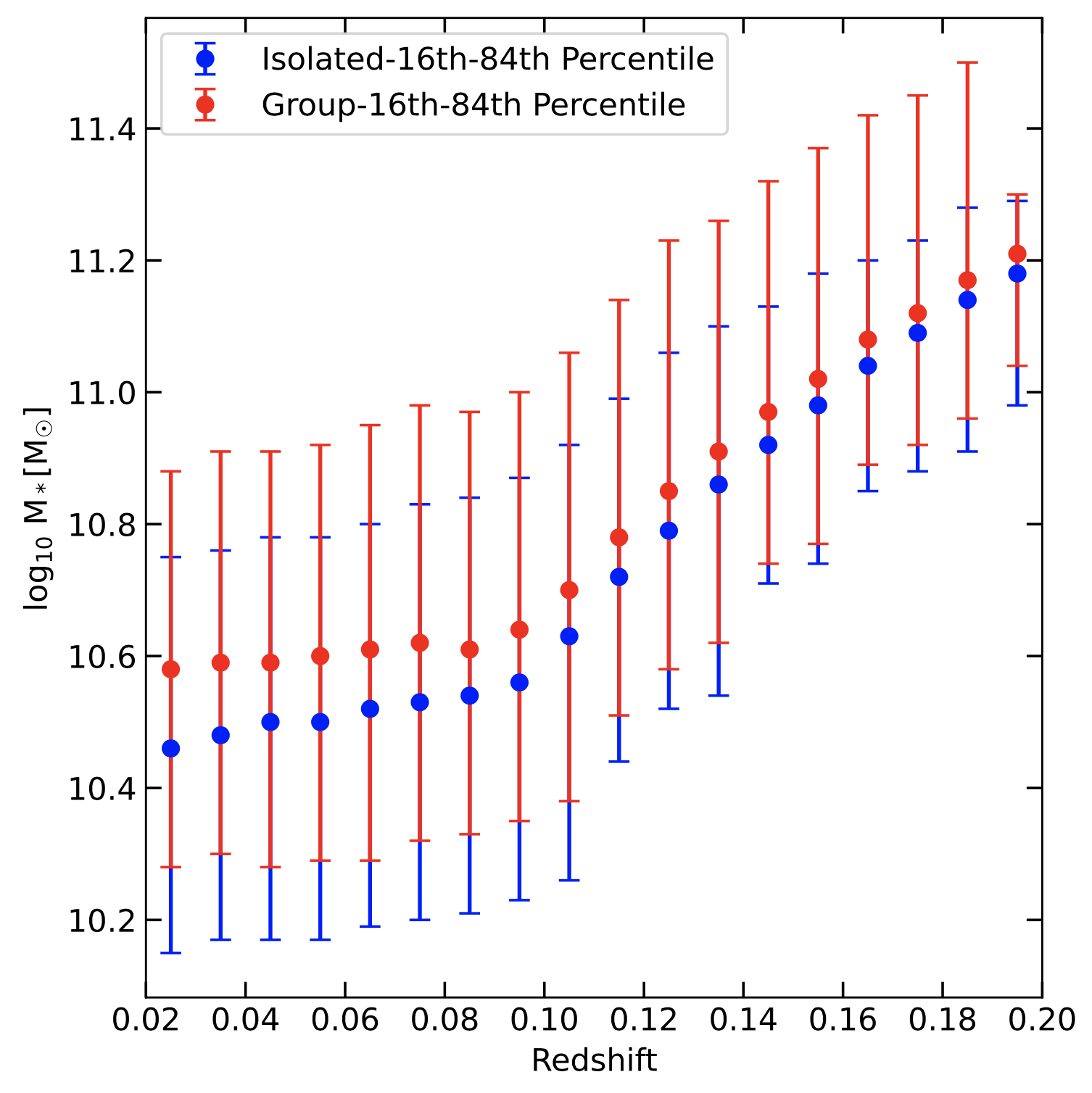}\hspace{0.1cm}
			\includegraphics[scale=0.21]{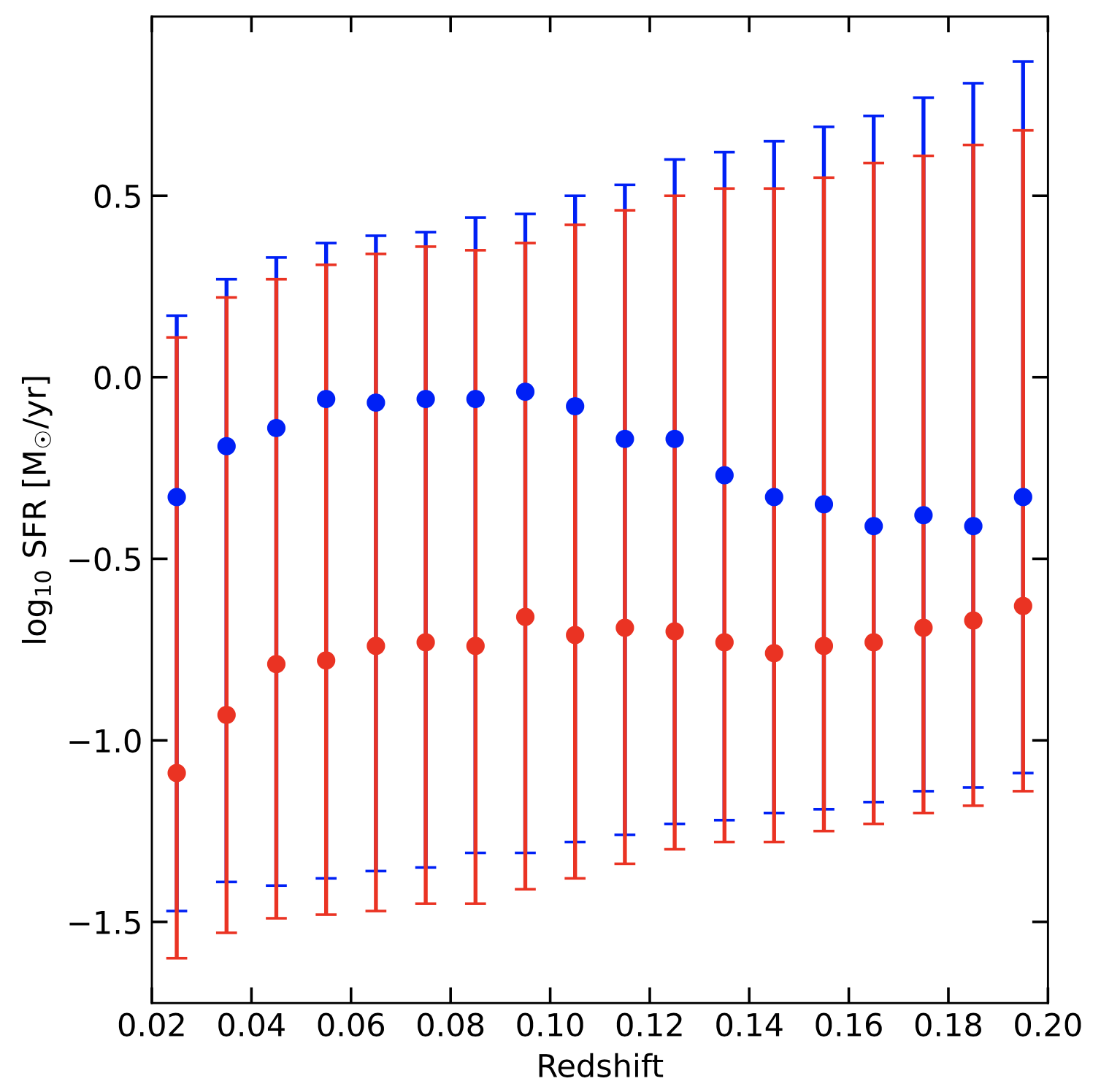}\hspace{0.1cm}
			\includegraphics[scale=0.21]{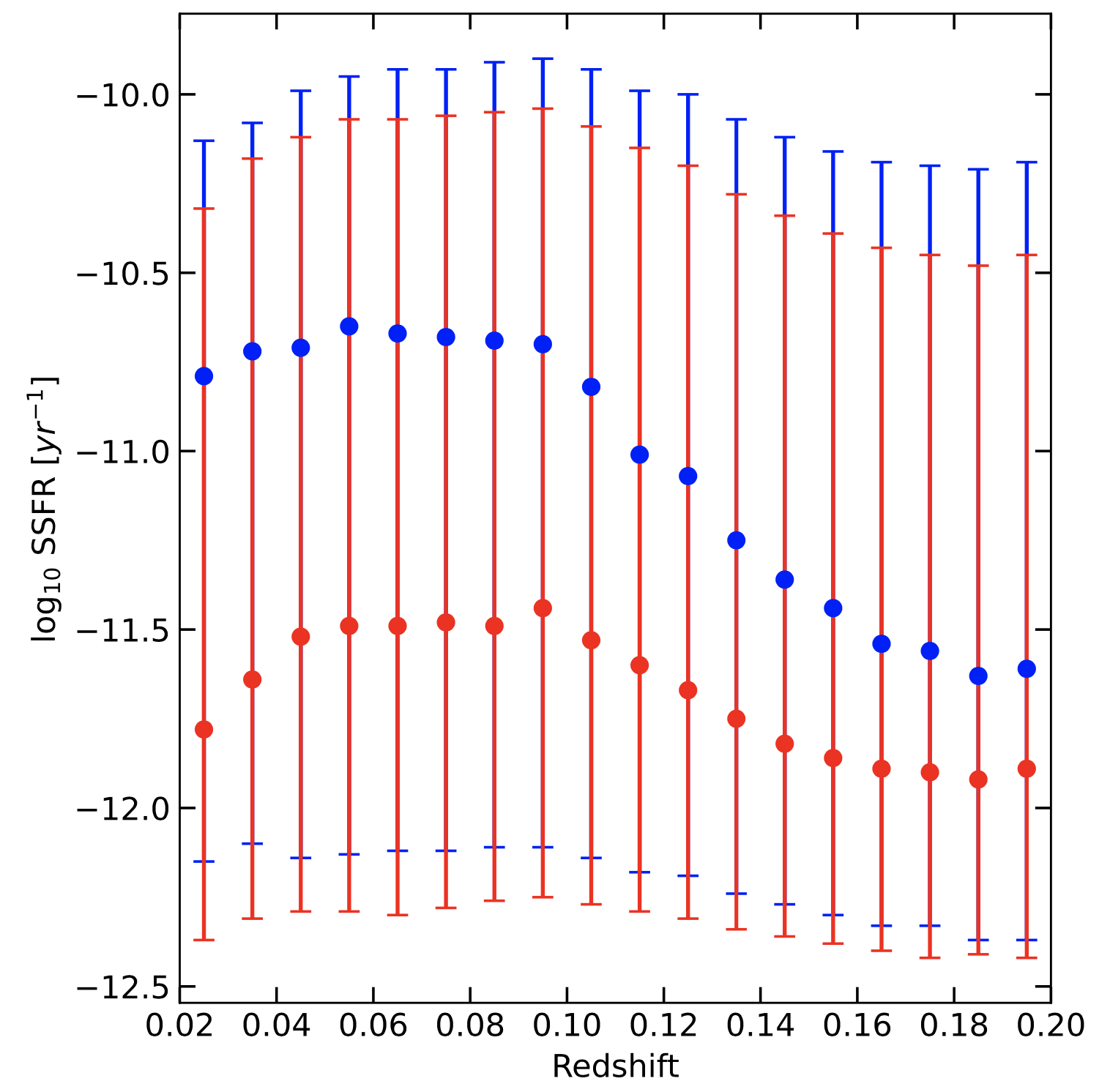}}
		\vspace{-0.2cm}
	 \caption{Variation of median stellar mass (left plot), SFR (middle plot), and 
				SSFR (right plot) with redshift for isolated galaxies (blue solid line) and 
				group galaxies (red solid line). The bars in each measurement indicate the 16th and 84th percentiles of the estimate.} 
			\label{zmasssfrssfr}
	\end{figure}

	As already mentioned, the SSFR is defined as the SFR per unit stellar
	mass, and 
	these quantities were derived by the MPA-JHU group. Figure 
	\ref{stellarmass} 
	shows the general distribution of stellar masses for the galaxies in each 
	redshift 
	bin within the range $0.02 \leq z \leq 0.2$, in both 
	isolated 
	and group environments. Table \ref{mp}, shows the average stellar mass, 
	SFR and SSFR 
	of isolated and group galaxies in each redshift bin.  As depicted in
	columns (2), (3) 
	of Table \ref{mp}, high-mass galaxies are consistently found in group 
	environments 
	across all redshift bins, while low-mass galaxies tend to preferentially inhabit in 
	isolated environments. The left plot of Figure \ref{zmasssfrssfr} indicates that the group 
	galaxies are more massive than the isolated galaxies across all redshift bins. This observation aligns with the findings of 
	\citet{kauffmann2004environmental} and  \citet{li2006dependence}, 
	suggesting that 
	high-mass galaxies are more found in the densest regions of 
	the Universe, 
	whereas low-mass galaxies are inclined to be located in low-density regions.
	\begin{figure}[!p]
		\centering
		\includegraphics[scale=0.15]{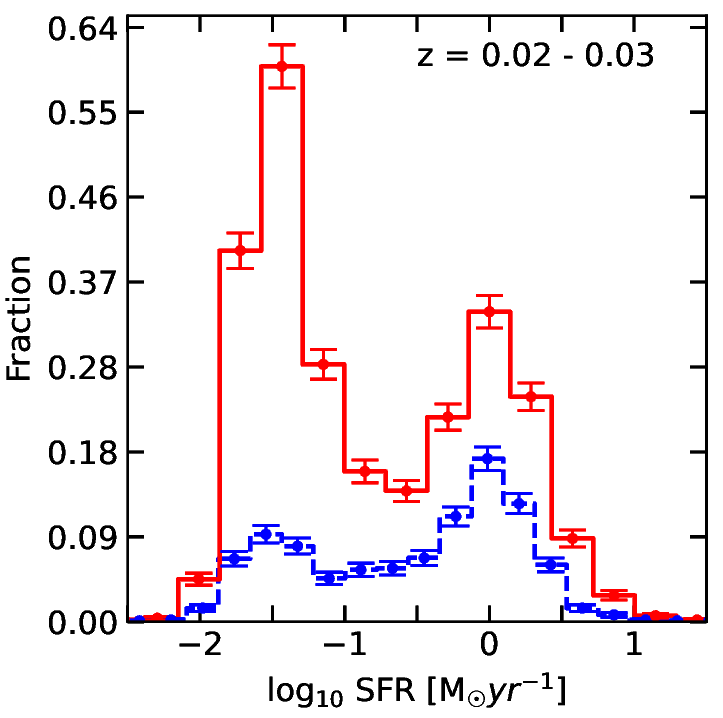}\hspace{0.5cm}
		\includegraphics[scale=0.15]{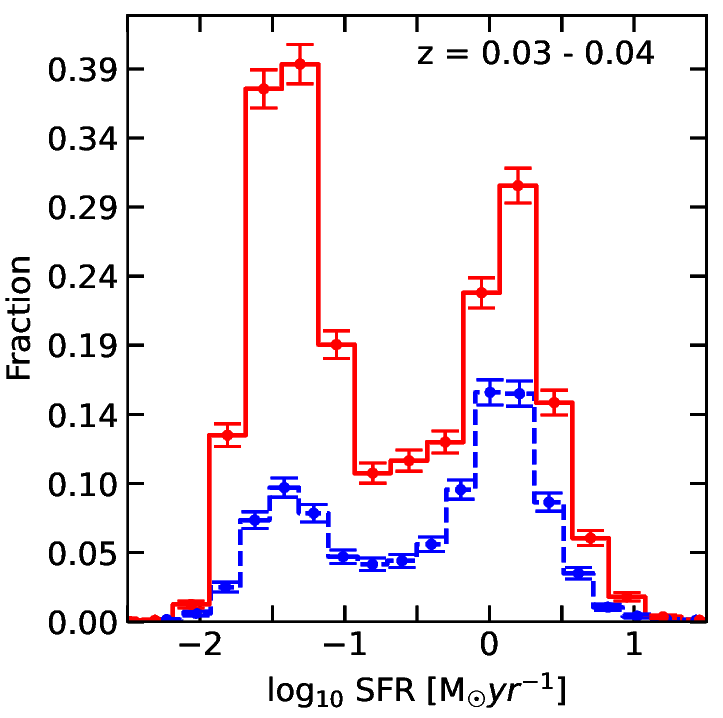}\hspace{0.5cm}
		\vspace{0.5cm}
		\includegraphics[scale=0.15]{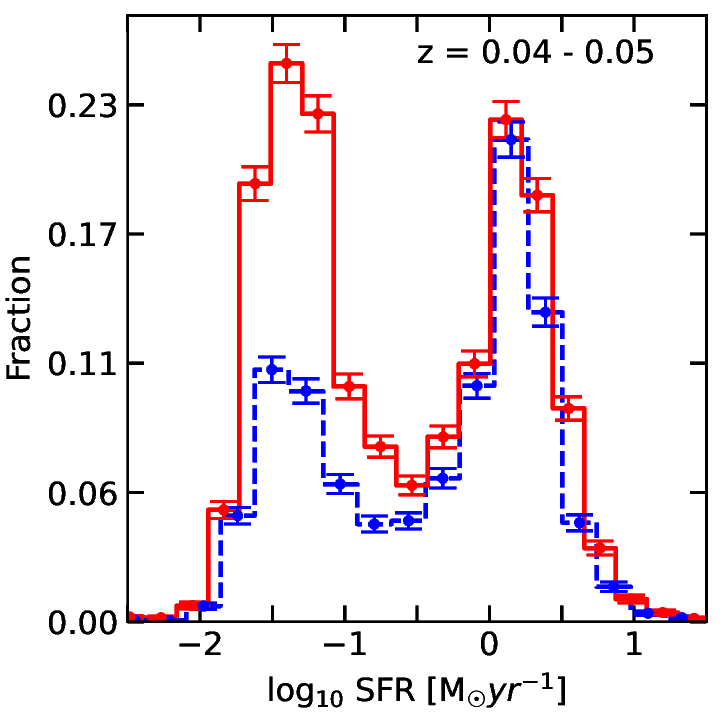}\hspace{0.5cm}
		\includegraphics[scale=0.15]{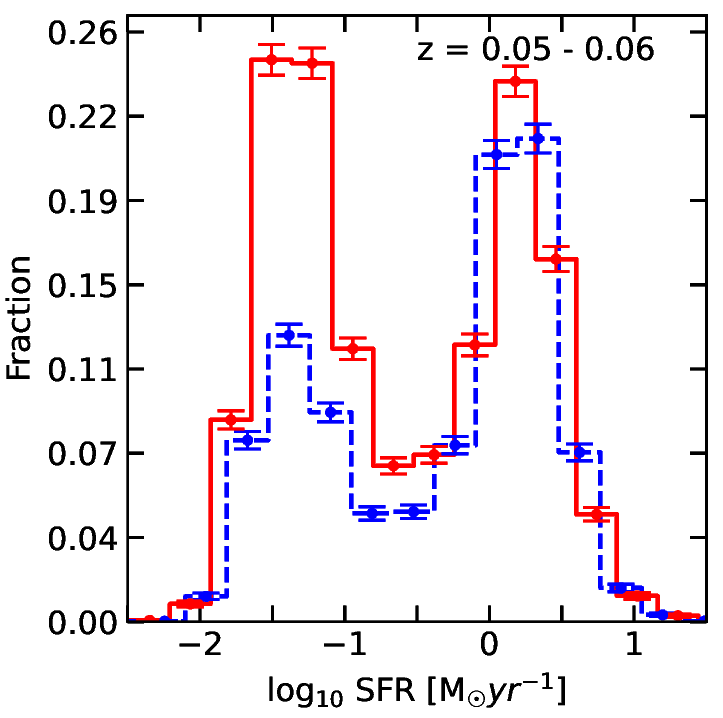}\hspace{0.5cm}
		\includegraphics[scale=0.15]{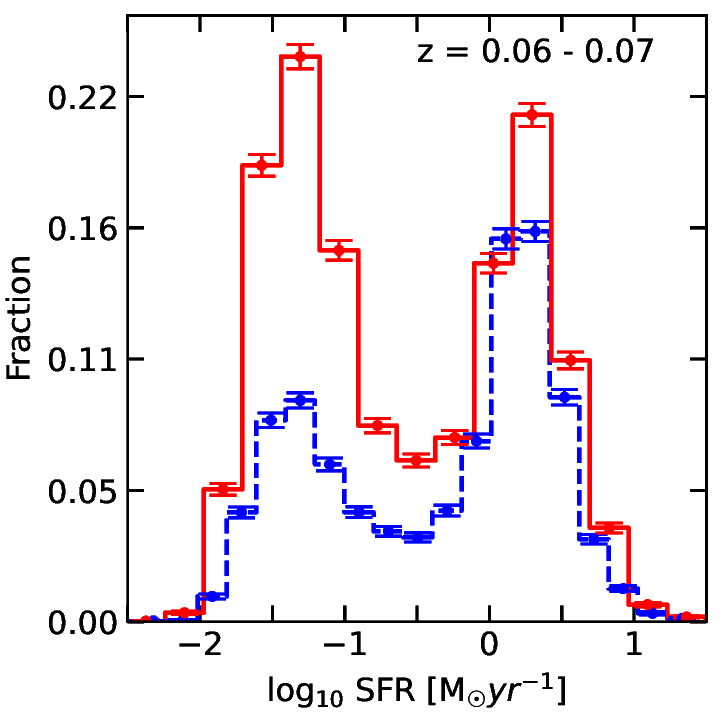}\hspace{0.5cm}
		\includegraphics[scale=0.15]{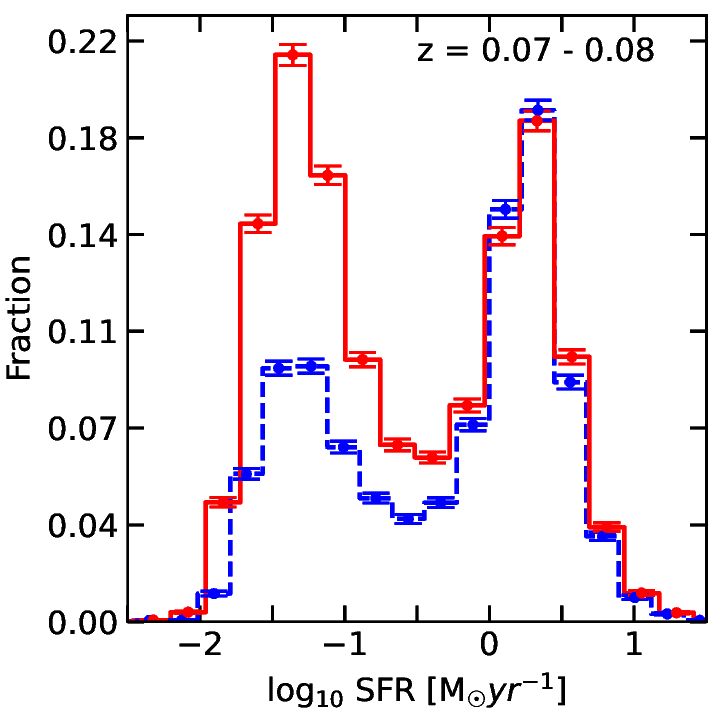}\hspace{0.5cm}
		\vspace{0.5cm}
		\includegraphics[scale=0.15]{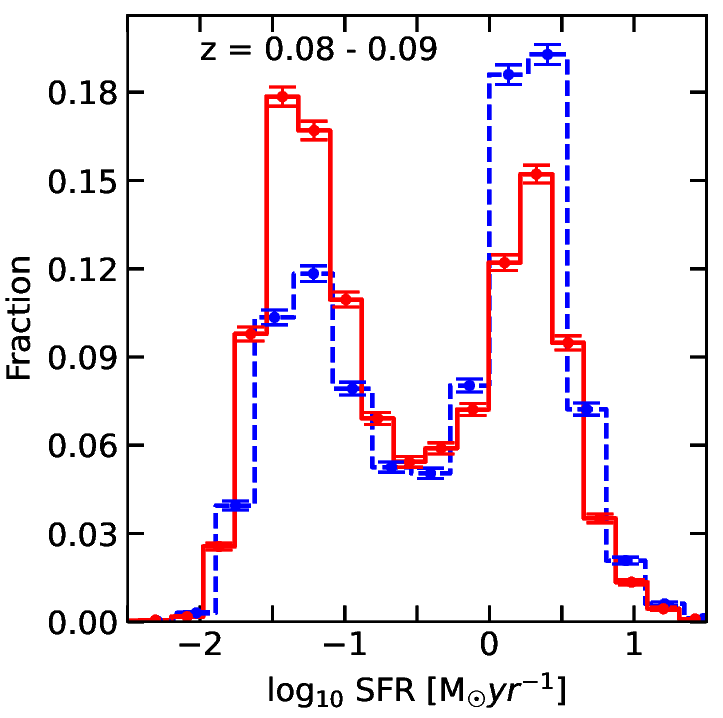}\hspace{0.5cm}
		\includegraphics[scale=0.15]{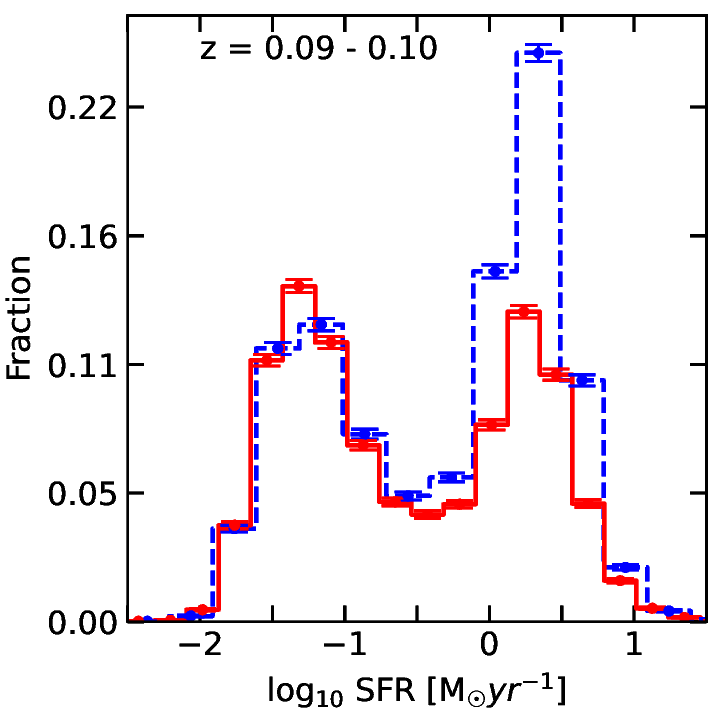}\hspace{0.5cm}
		\includegraphics[scale=0.15]{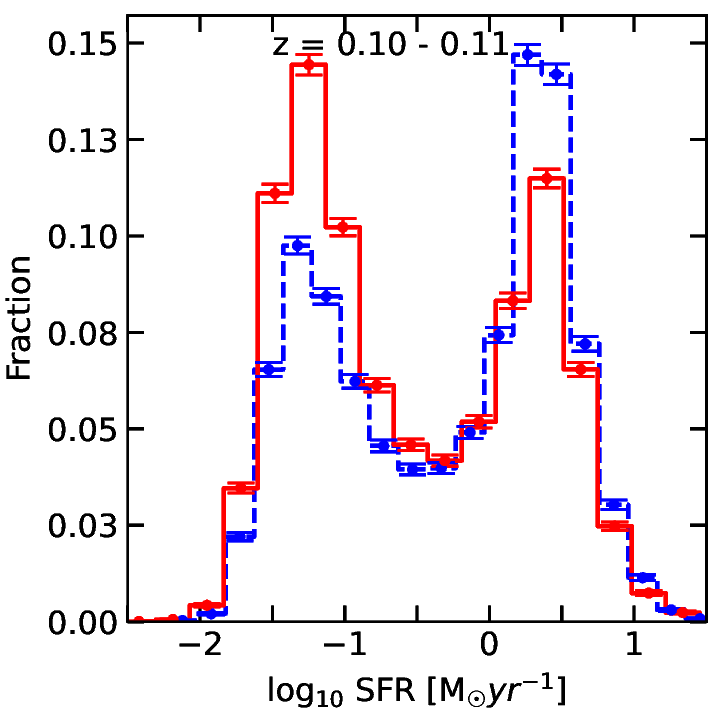}\hspace{0.5cm}
		\includegraphics[scale=0.15]{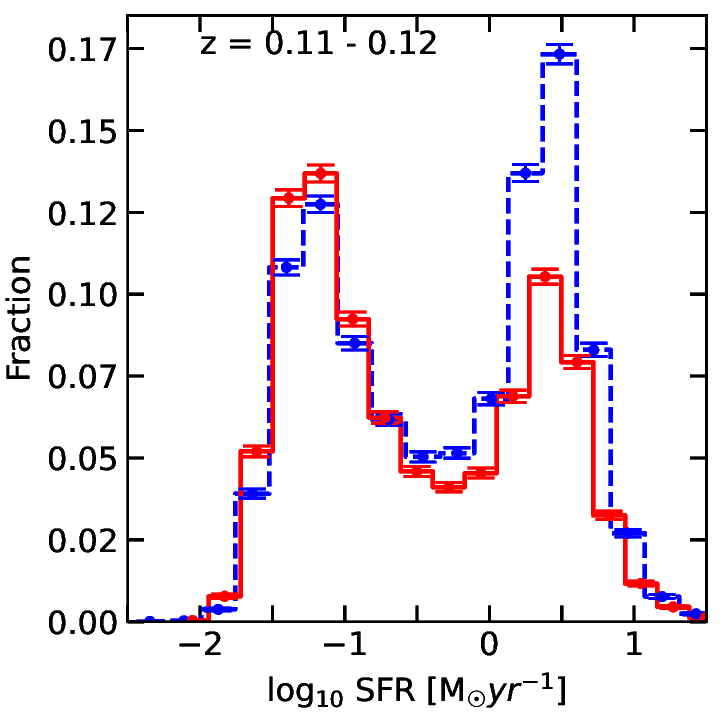}\hspace{0.5cm}
		\vspace{0.5cm}
		\includegraphics[scale=0.15]{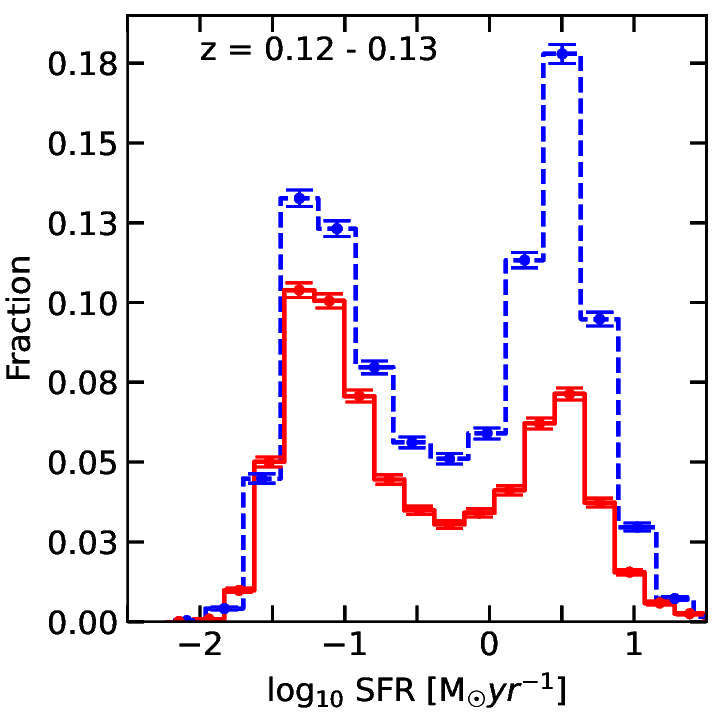}\hspace{0.5cm}
		\includegraphics[scale=0.15]{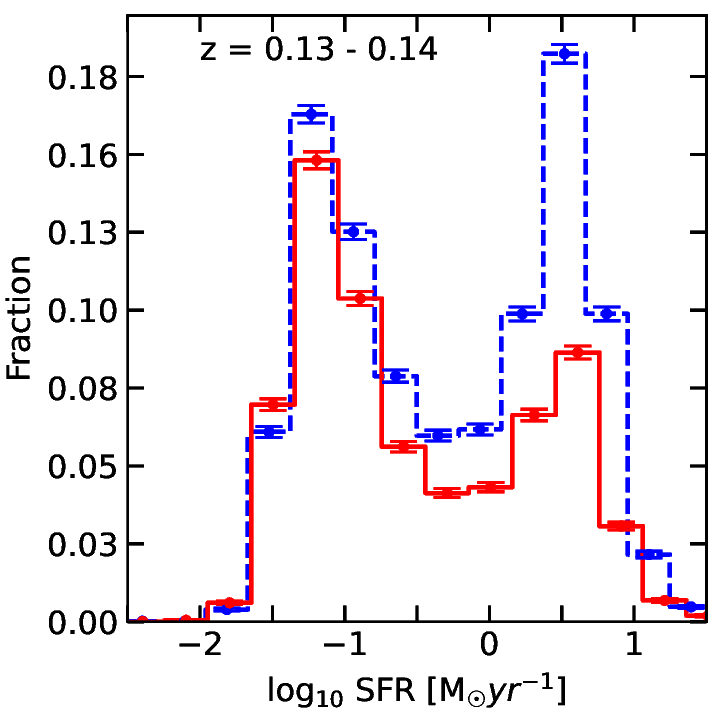}\hspace{0.5cm}
		\includegraphics[scale=0.15]{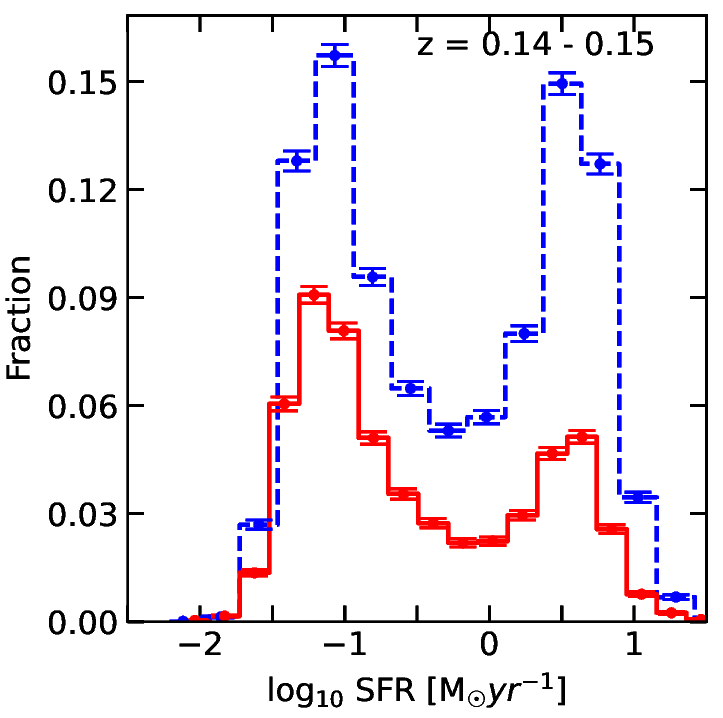}\hspace{0.5cm}
		\includegraphics[scale=0.15]{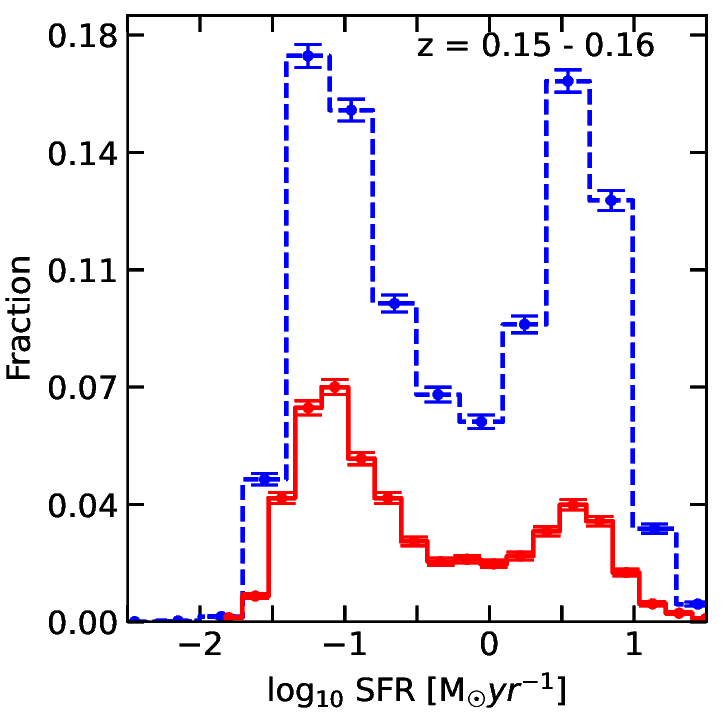}\hspace{0.5cm}
		\vspace{0.5cm}
		\includegraphics[scale=0.15]{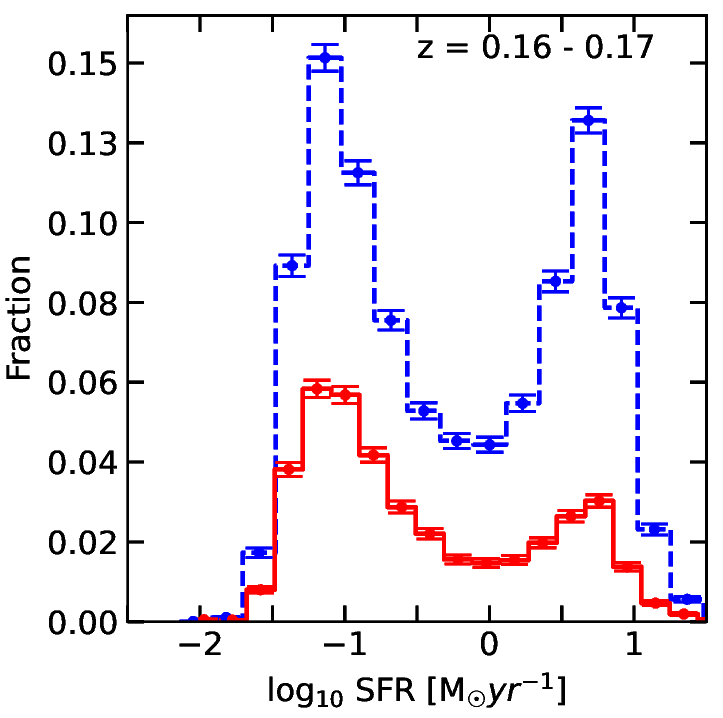}\hspace{0.5cm}
		\includegraphics[scale=0.15]{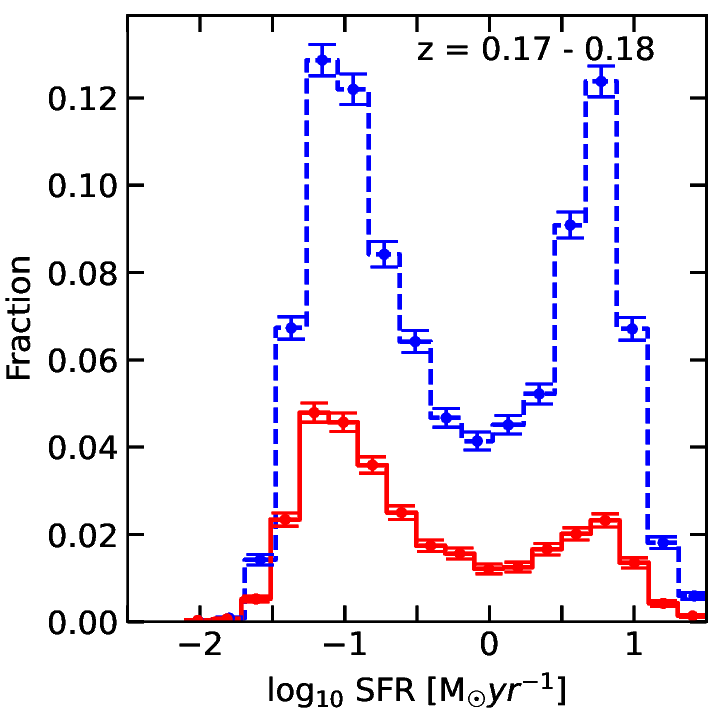}\hspace{0.5cm}
		\includegraphics[scale=0.15]{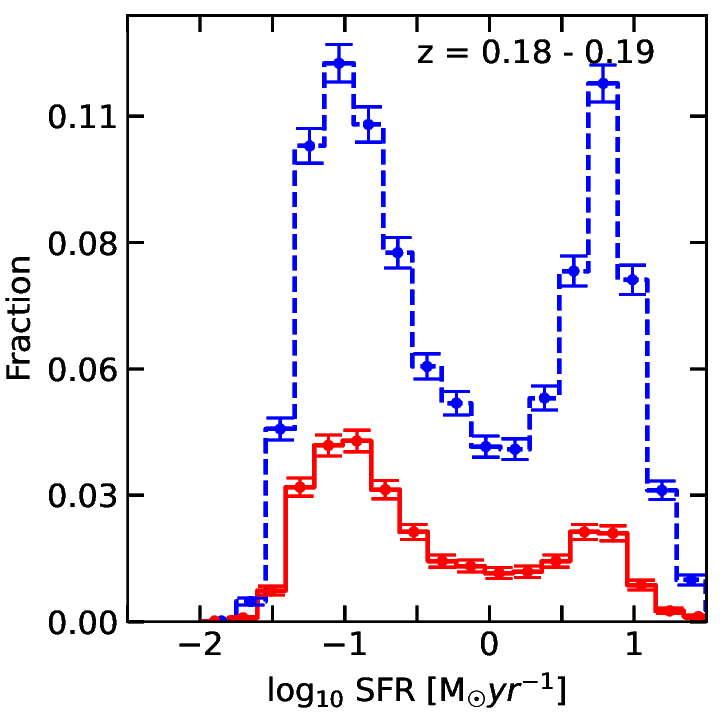}\hspace{0.5cm}
		\includegraphics[scale=0.15]{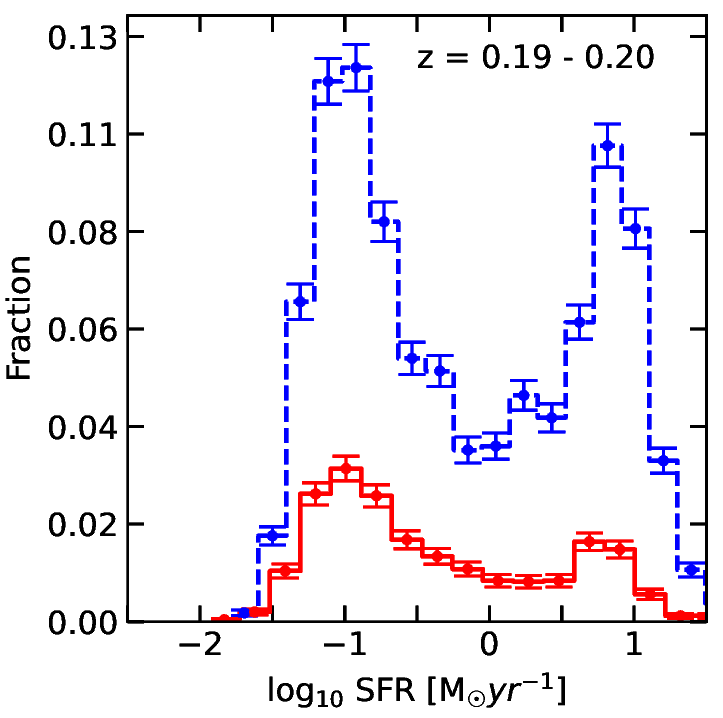}\hspace{0.5cm}
		
		\caption{Star formation rate distributions for isolated galaxies (blue dotted line) 
			and group galaxies (red solid line). The error bars are 1 $\sigma$ Poissonian errors.} 
		\label{sfr}
		\vspace{1cm}
	\end{figure}
	
\begin{figure}[!p]
		\centering
		\includegraphics[scale=0.15]{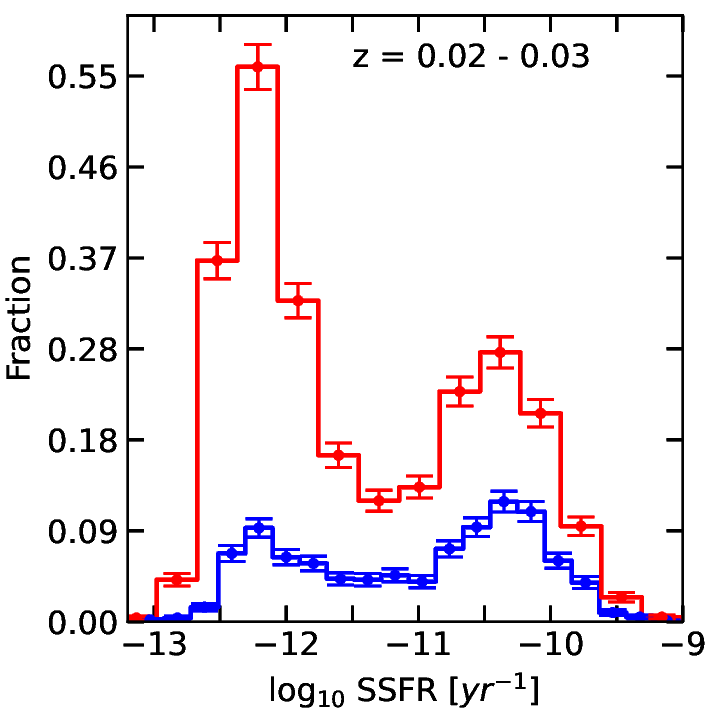}\hspace{0.5cm}
		\includegraphics[scale=0.15]{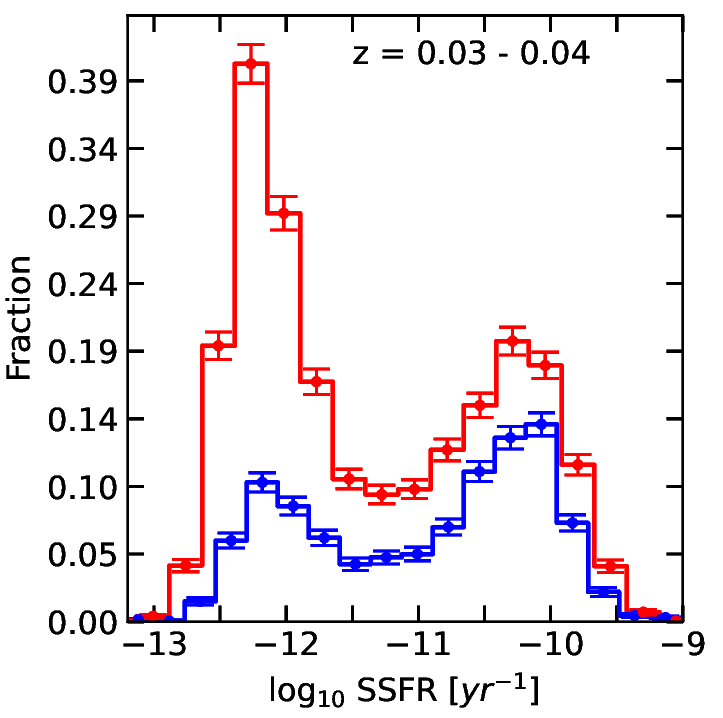}\hspace{0.5cm}
		\vspace{0.5cm}
		\includegraphics[scale=0.15]{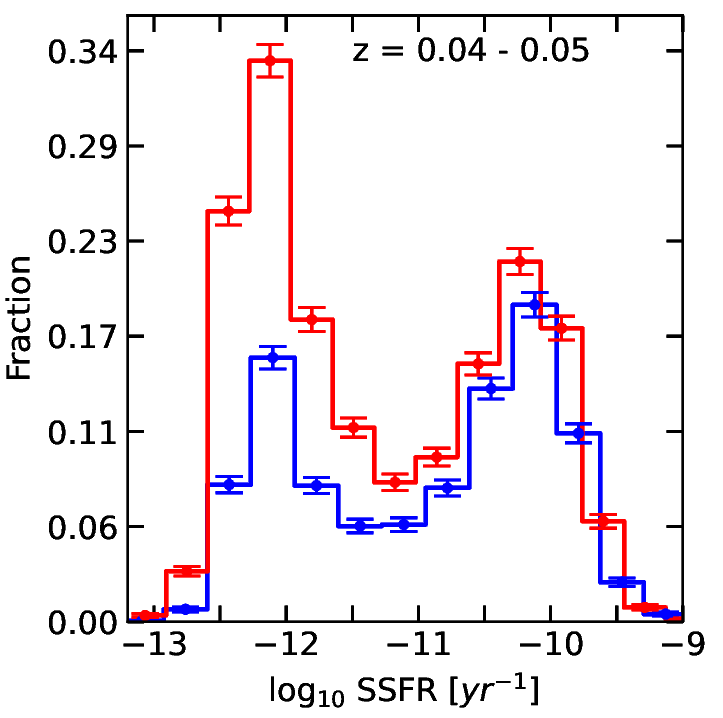}\hspace{0.5cm}
		\includegraphics[scale=0.15]{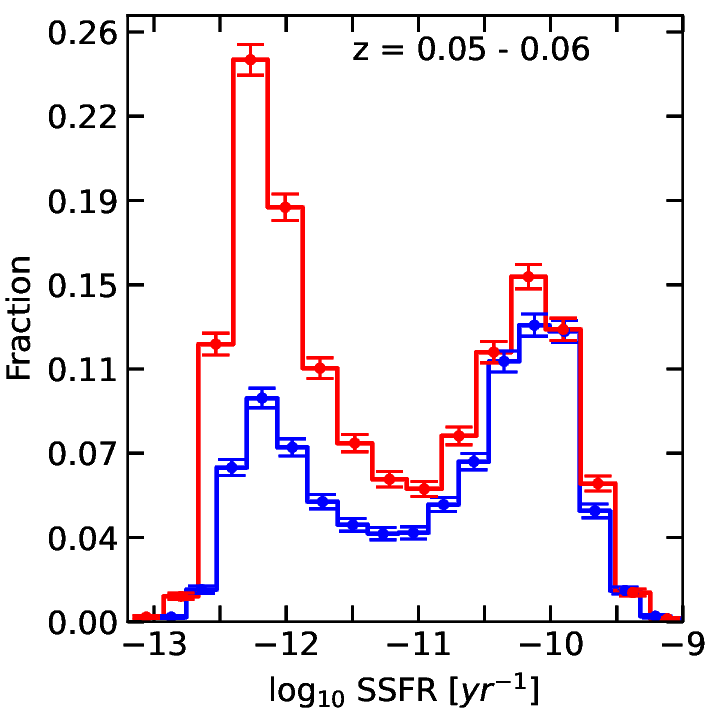}\hspace{0.5cm}
		\includegraphics[scale=0.15]{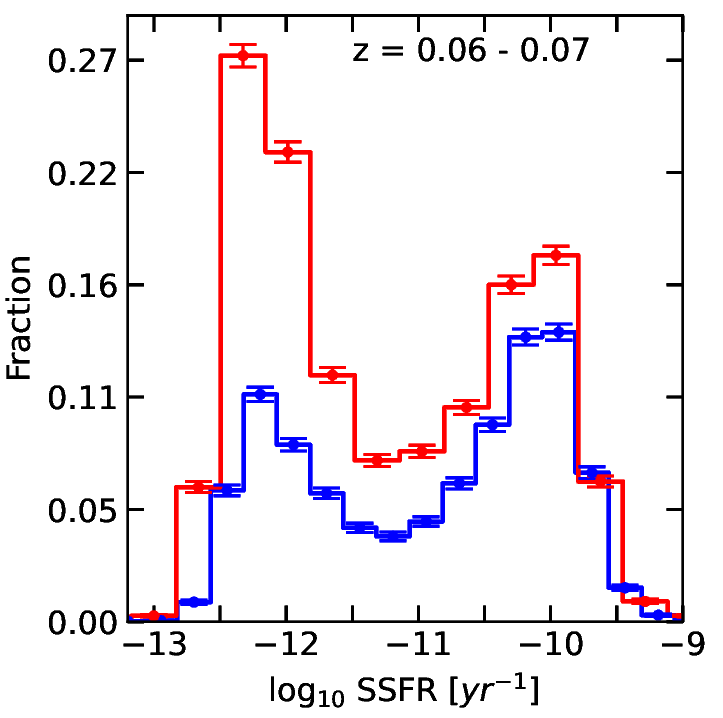}\hspace{0.5cm}
		\includegraphics[scale=0.15]{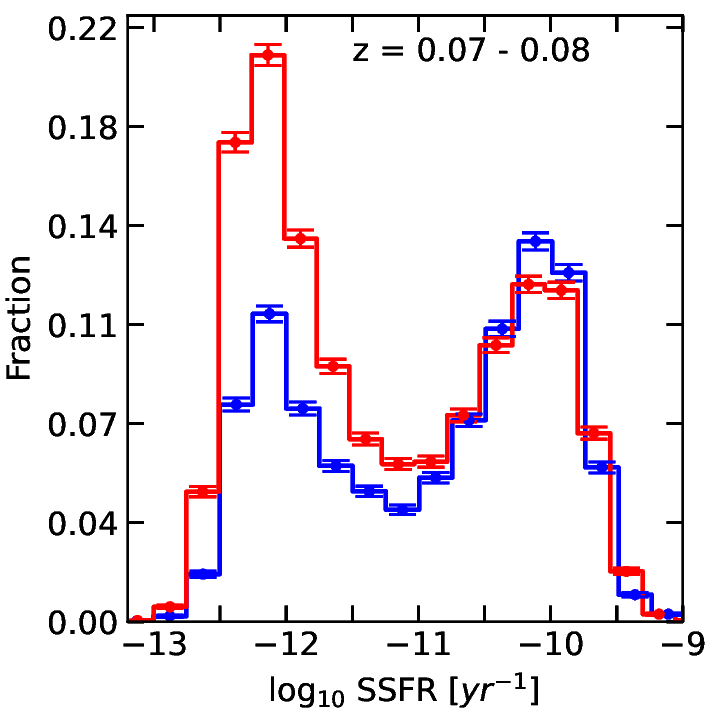}\hspace{0.5cm}
		\vspace{0.5cm}
		\includegraphics[scale=0.15]{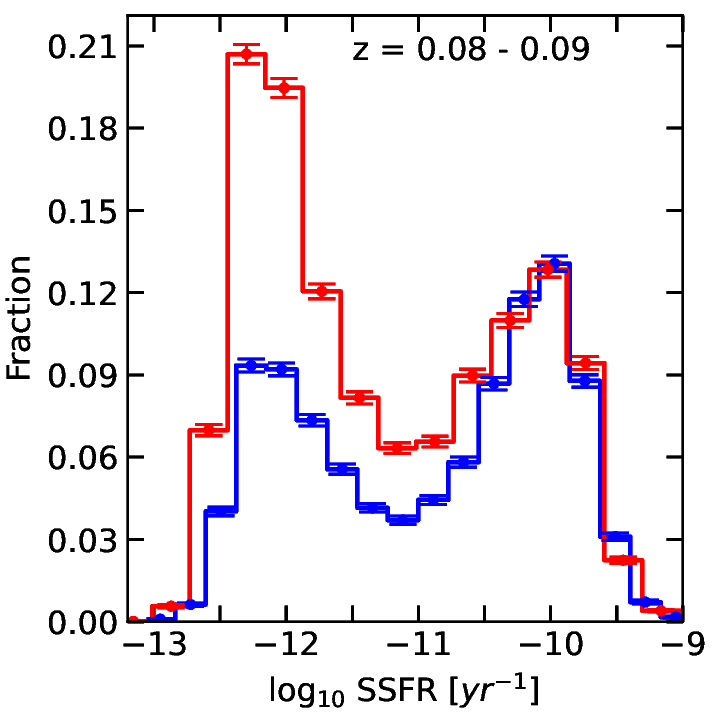}\hspace{0.5cm}
		\includegraphics[scale=0.15]{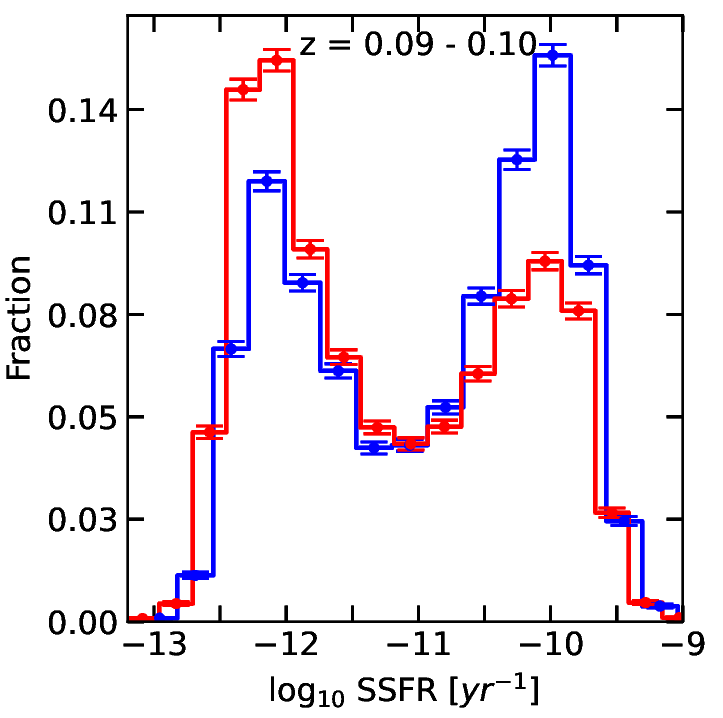}\hspace{0.5cm}
		\includegraphics[scale=0.15]{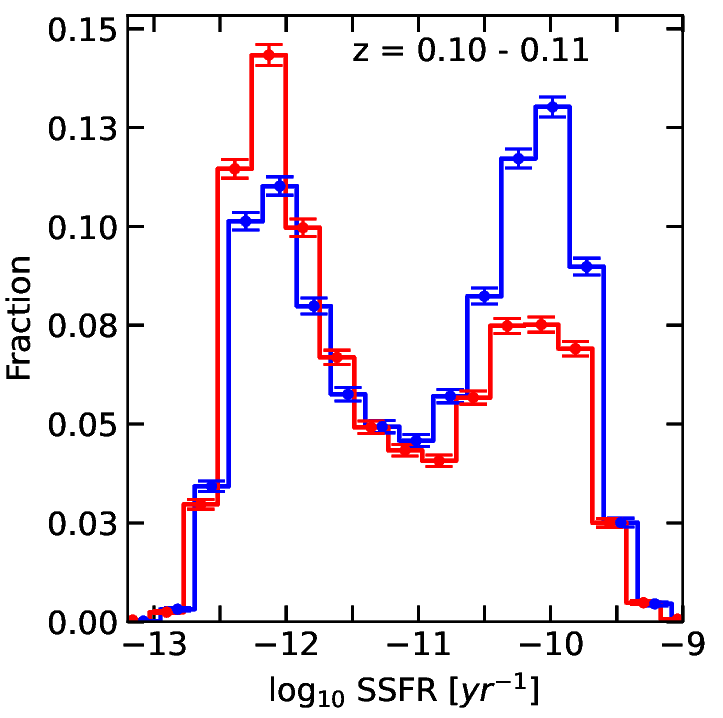}\hspace{0.5cm}
		\includegraphics[scale=0.15]{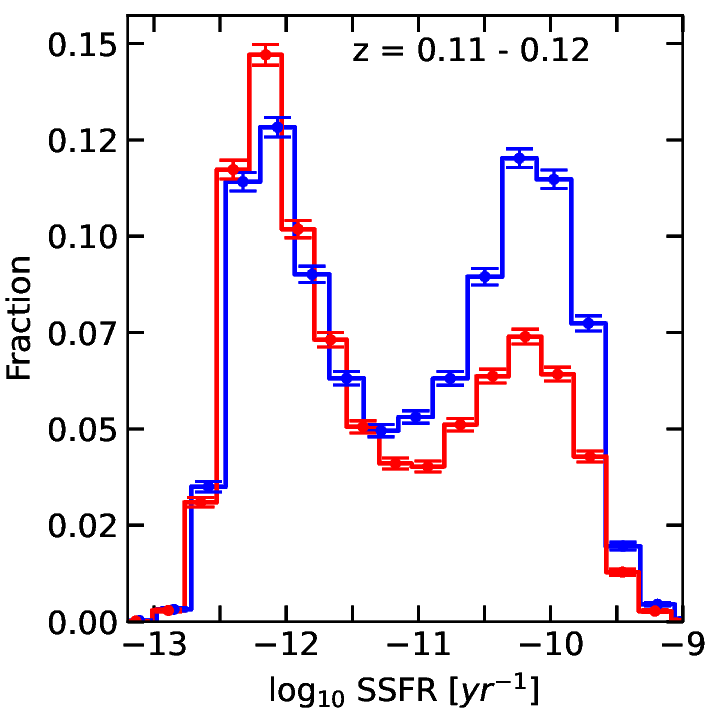}\hspace{0.5cm}
		\vspace{0.5cm}
		\includegraphics[scale=0.15]{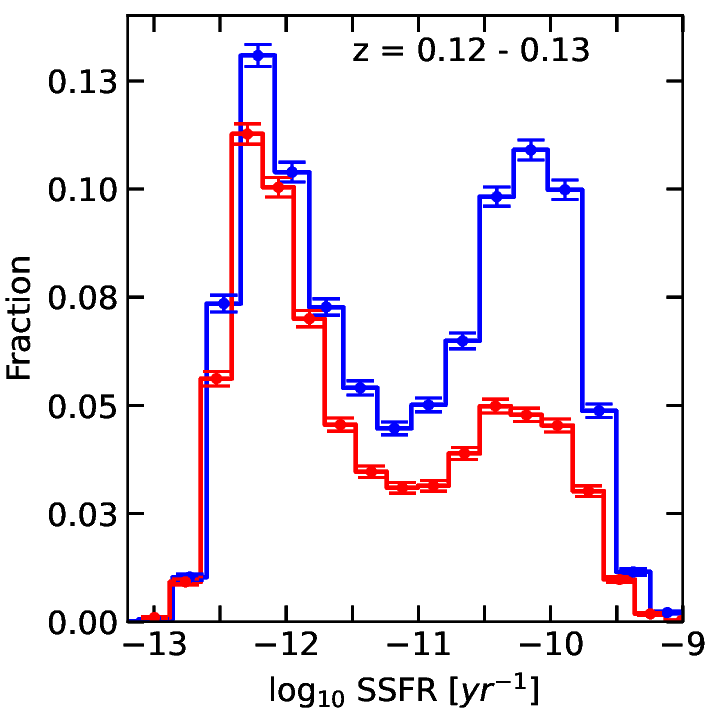}\hspace{0.5cm}
		\includegraphics[scale=0.15]{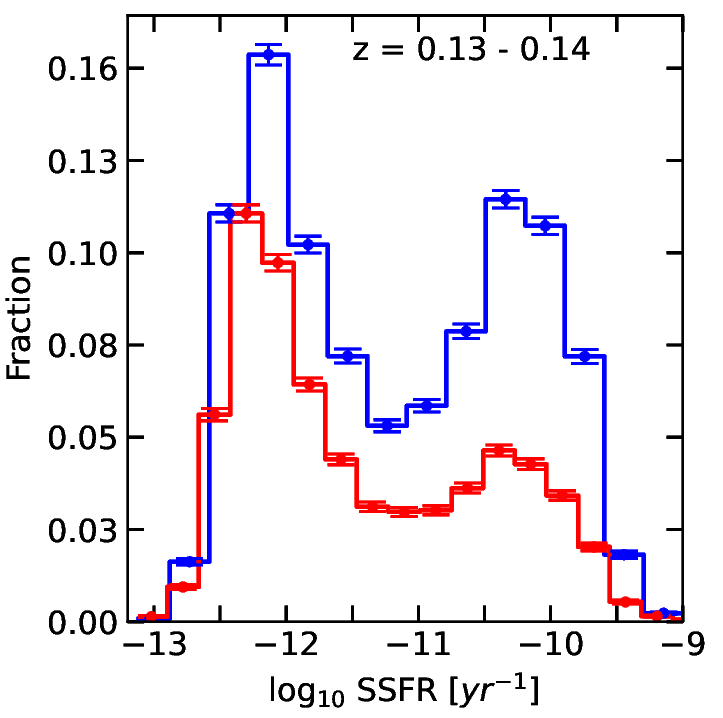}\hspace{0.5cm}
		\includegraphics[scale=0.15]{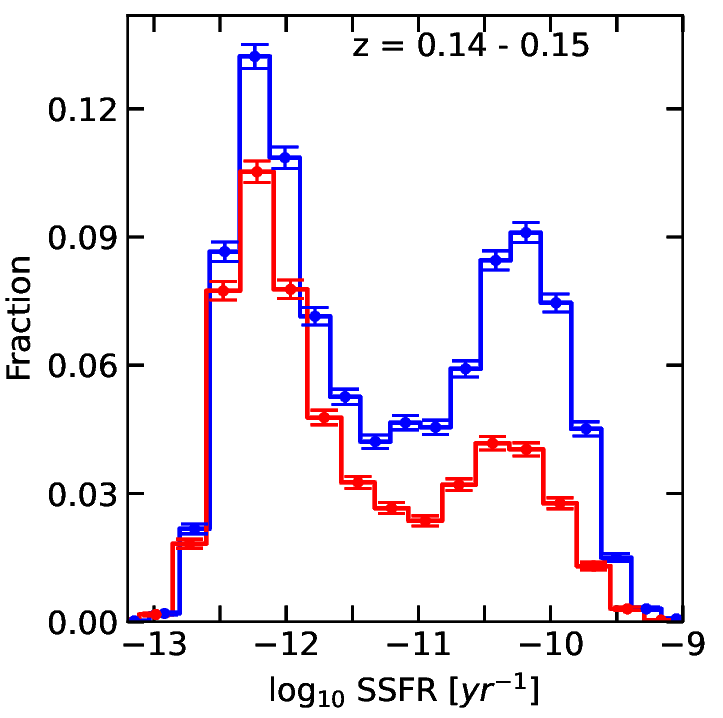}\hspace{0.5cm}
		\includegraphics[scale=0.15]{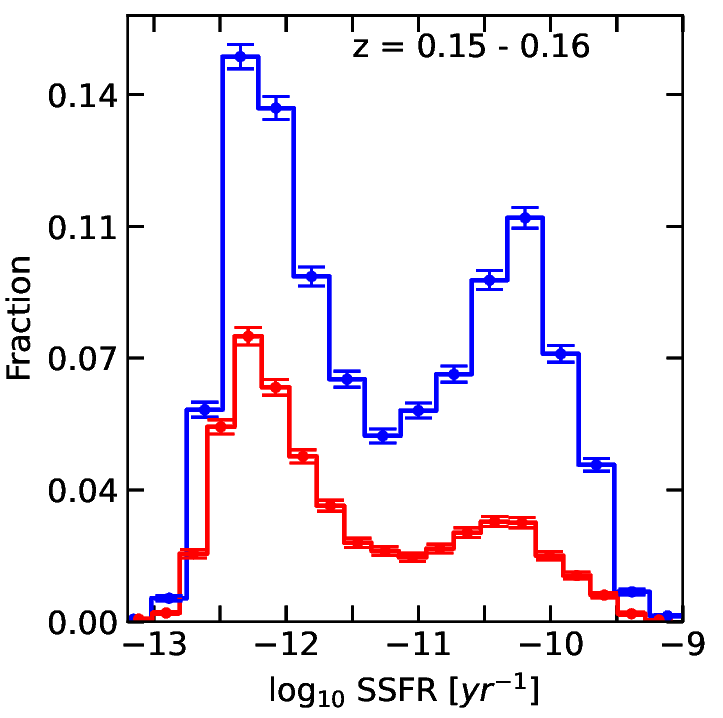}\hspace{0.5cm}
		\vspace{0.5cm}
		\includegraphics[scale=0.15]{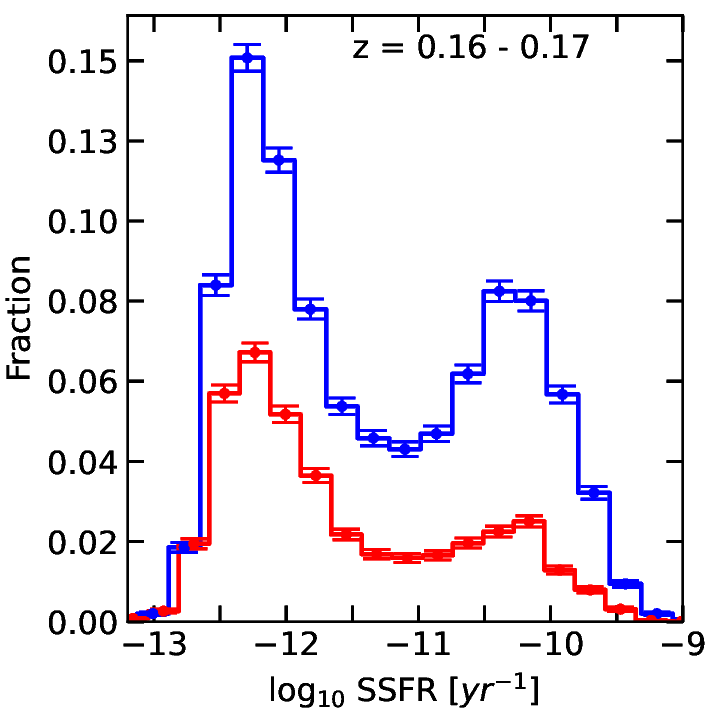}\hspace{0.5cm}
		\includegraphics[scale=0.15]{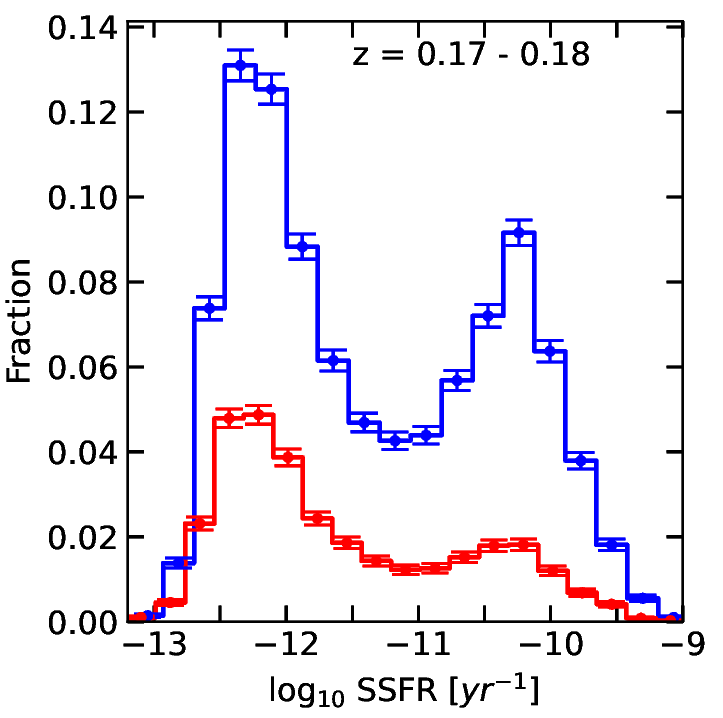}\hspace{0.5cm}
		\includegraphics[scale=0.15]{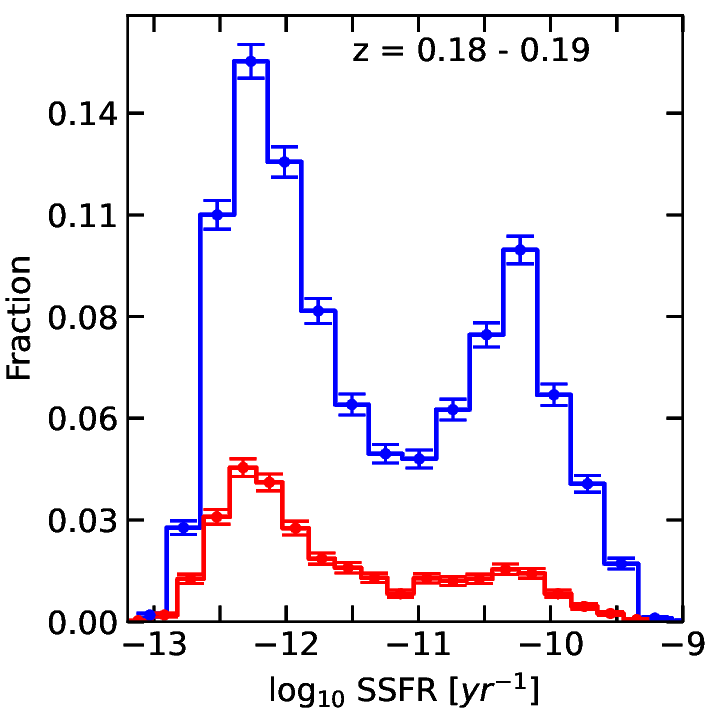}\hspace{0.5cm}
		\includegraphics[scale=0.15]{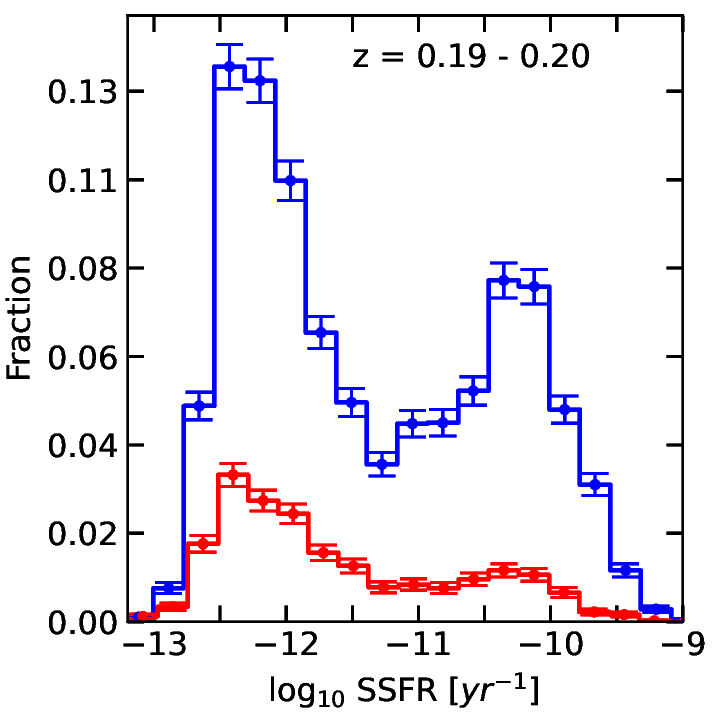}\hspace{0.5cm}
		\caption{Specific star formation rate distribution for isolated galaxies (blue dotted line) 
			and group galaxies (red solid line). The error bars are 1 $\sigma$ Poissonian errors.} 
		\label{ssfr}
		\vspace{1cm}
	\end{figure}

	Figures \ref{sfr} and \ref{ssfr} show the general distributions of SFR and SSFR,
	respectively 
	within each redshift bins for the redshift range of $0.02 \leq z \leq 0.2$ in both 
	isolated and 
	group environments. As depicted in columns (4), (5) of Table \ref{mp}, galaxies in isolated environments 
	tend to have 
	higher SFR than in isolated environments. Similarly, from  
	columns (6), 
	(7) of Table \ref{mp}, it is also evident that galaxies 
	exhibit low 
	SSFR in group environments than isolated environment across all redshift bins compared to 
	isolated environments. 
	The middle plot of Figure \ref{zmasssfrssfr} indicates 
	that group 
	galaxies have lower SFR than isolated galaxies, similarly  the right plot of Figure \ref{zmasssfrssfr} 
	indicates that 
	group galaxies possess higher SSFR than isolated galaxies. From the left plot of Figure \ref{zmasssfrssfr}, 
	it is observed 
	that the stellar mass increases with the increase in redshift having a steep slope at $z \gtrsim 12$. 
	Again from 
	Figure \ref{zmasssfrssfr} it is observed that the median difference between isolated and group galaxies 
	for stellar mass, 
	SFR and SSFR start to decreases at $z \gtrsim 12$. The observed consistency of stellar mass, SFR and SSFR in 
	isolated and 
	group environments supports the perspective that galaxies generally 
	display 
	low stellar mass, high SFR and SSFR in low-density regions, and high stellar mass, low SFR and SSFR in 
	high-density 
	regions of the Universe, which implies that high-density environments tend 
	to inhibit 
	the process of star formation indicated by many possible mechanisms 
	such as 
	supernova explosion, magnetic fields, stellar winds, and radiation pressure \cite{lewis2002df, gomez2003galaxy, tanaka2004environmental,elbaz2007reversal, cooper2008deep2, patel2009dependence}.
	Moreover, 
	the observation from Figures \ref{stellarmass}, \ref{sfr}, and 
	\ref{ssfr} 
	indicate that at high redshift bins $(z \gtrsim 0.12)$, the sample in 
	isolated 
	environments has a higher proportion of galaxies, whereas, at low 
	redshift 
	bins $(z \lesssim 0.1)$, the sample in group environments has a higher 
	proportion 
	of galaxies, for intermediate redshift $(0.1 \lesssim z \lesssim 0.12)$ an 
	approximately equal proportional of galaxies is observed.
	The Kolmogorov-Smirnov (KS) as detailed in Ref.\ \cite{hodges1958significance,harari2009kolmogorov}, and Anderson–Darling (AD) as refer to Ref.\ \cite{babu2006astrostatistics} statistical tests
	serve as 
	a quantitative comparison, assessing the degree of 
	similarity 
	or difference between two independent distributions by calculating the KS, AD statistic and probability values. 
	The KS, AD statistics and probability 
	values indicates the likelihood that the two distributions are derived from the 
	same parent 
	distribution. When the KS, AD statistics values is closer to zero indicates that the two distributions are similar, again a high probability suggests a high likelihood of the two distributions 
		sharing a 
		common parent, while a low probability implies that the distributions are different. We carried out a two-sample KS, and AD tests between isolated and group galaxies 
		across all redshift bins for stellar mass, SFR and SSFR. The calculated  KS statistics,
		 listed in columns (2), (3) and (4) of Table \ref{st} are between $\sim 0.10$ and $\sim 0.18$, while p-values between $\sim 2.41 \times 10^{-101}$ 
		and $\sim 6.58\times 10^{-12}$ 
		for stellar mass, between $\sim 0.11$ and $\sim 0.21$, while p-values between $\sim 6.63 \times 10^{-191}$ and $\sim 4.86 \times 10^{-14}$ 
		for SFR, 
		similarly between $\sim 0.12$ and $\sim 0.22$, while p-values between $\sim 6.39 \times 10^{-197}$ and $\sim 2.46 \times 10^{-16}$ for SSFR, respectively. The KS statistics are all greater than zero and the p-values are 
		notably much less than 0.05 $(5\%)$, a standard threshold in statistical analysis. Again the calculated  AD statistics, listed in columns (5), (6)  and (7) of Table \ref{st}, are between $\sim 36.29$ and $\sim 448.83$, while p-values are $\sim \num{1e05}$
		for stellar mass, between $\sim 30.38$ and $\sim 472.25$, while p-values are $\num{1e05}$
		for SFR, 
		similarly between $\sim 33.95$ and $\sim 704.41$, while p-values are $\sim \num{1e05}$ for SSFR, respectively. The AD statistics are all far greater than zero and the p-values are 
		notably much less than 0.05 $(5\%)$, a standard threshold in statistical analysis. This outcome 
	suggests that the two independent distributions in each of Figures \ref{stellarmass}, 
	\ref{sfr}, and 
	\ref{ssfr} are significantly different, reinforcing that a strong 
	correlation 
	exists between the environment, stellar mass, SFR, and SSFR. These 
	findings 
	support the assertion of a robust environmental dependence of stellar mass, 
	SFR, and 
	SSFR in all redshift bins within the sample. Consequently, there 
	is substantial 
	confidence in accepting this conclusion. Again from Table \ref{st} 
	it is observed 
	that the dependence of stellar mass, SFR and SSFR on environment is weak at 
	higher redshift 
	when compared to low redshift based on KS-probabilities the strong dependence of all the properties is found at 
	the range of 
	$0.07\leq z\leq 0.09$ while the weak dependence is found at $0.19\leq z\leq 0.2$. Again from column 
	(3), (4) and (5) of Table \ref{st} the SSFR strongly depend on environment 
	across all 
	redshift bins with  $6.39\times 10^{-197} \leq P\leq 2.46\times 10^{-16}$ followed by 
	SFR with 
	$6.63\times 10^{-191} \leq P\leq 4.86\times 10^{-14}$ when compared with 
	stellar mass 
	with P-value within the range $2.41\times 10^{-101} \leq P\leq 6.58\times 10^{-12}$.
	
	%\onecolumn
	\begin{table}[!h]
		\centering
	    \caption{Median stellar mass (M$\star$), median SFR and median SSFR for isolated and
				group galaxies within the redshift bin size of $\Delta z = 0.01$. The 
				associated uncertainty is the 16th and 84th percentiles of the estimate.}
			\vspace{8pt}
			\setlength{\tabcolsep}{0.6pc}
			\scalebox{0.97}{
				\begin{tabular}{cccccccc}
					\toprule
					\toprule
					Redshift  & Galaxy number&\multicolumn{2}{c}{log$_{10}$ M$\star[M_{\odot}]$} & \multicolumn{2}{c}{log$_{10}$ SFR$[M_{\odot} /yr$]} & \multicolumn{2}{c}{log$_{10}$ SSFR$[yr^{-1}]$} \\
					\cmidrule(lr){3-4} \cmidrule(lr){5-6} \cmidrule(lr){7-8}  % Adding partial midrules
					&  &Isolated & Group & Isolated & Group & Isolated & Group \\
					{(1)} &(2)&{(3)}&{(4)}&{(5)}&{(6)}&{(7)}&{(8)}\\
					\midrule
					$0.02\leq z<0.03$ & $3881$&$10.46^{+0.29}_{-0.31} $ & $10.58^{+0.31}_{-0.30}$ & $ -0.33^{+0.50} _{-1.14}$& $-1.09 ^{+1.20} _{-0.51}$  & $-10.79 ^{+0.66} _{-1.36}$  & $-11.78 ^{+1.46} _{-0.59}$ \\
					$0.03\leq z<0.04$ &$6651$&$10.48^{+0.28} _{-0.30} $& $10.59 ^{+0.32} _{-0.31}$  & $ -0.19 ^{+0.46 } _{-1.20}$ &  $-0.93^{+1.15} _{-0.60}$ & $-10.72 ^{+0.64} _{-1.38}$  & $-11.64^{+1.46} _{-0.67}$  \\
					$0.04\leq z<0.05$ &$9507$& $10.50^{+0.31} _{-0.34}$ &$10.59 ^{+0.32} _{-0.31}$ & $ -0.14 ^{+0.47} _{-1.26}$&  $-0.79 ^{+1.06} _{-0.70}$ & $-10.71 ^{+0.72} _{-1.43}$  & $-11.52 ^{+1.40} _{-0.77}$ \\
					$0.05\leq z<0.06$ &$13105$& $10.50 ^{+0.32} _{-0.33}$& $10.60^{+0.33} _{-0.32}$   & $ -0.06^{+0.43} _{-1.32}$&  $-0.78 ^{+1.09} _{-0.70}$ & $-10.65 ^{+0.70} _{-1.48}$  & $-11.49^{+1.42} _{-0.80}$ \\
					$0.06\leq z<0.07$ &$21851$&  $10.52 ^{+0.32} _{-0.34}$& $10.61^{+0.33} _{-0.31}$  & $ -0.07^{+0.46} _{-1.29}$&  $-0.74^{+1.08} _{-0.73}$ & $-10.67 ^{+0.74 } _{-1.45}$  & $-11.49^{+1.42} _{-0.81}$ \\
					$0.07\leq z<0.08$ &$32717$ & $10.53 ^{+0.31} _{-0.33} $& $10.62 ^{+0.32} _{-0.31}$  & $ -0.06^{+0.46} _{-1.29}$&  $-0.73^{+1.09} _{-0.72}$ & $-10.68^{+0.75} _{-1.44}$  & $-11.48^{+1.42} _{-0.80}$ \\
					$0.08\leq z<0.09$ &$37742$&  $10.54 ^{+0.33} _{-0.33} $& $10.61^{+0.33} _{-0.30}$  & $ -0.06^{+0.50} _{-1.25}$ & $-0.74^{+1.09} _{-0.71}$ & $-10.69^{+0.78} _{-1.42}$ & $-11.49 ^{+1.44} _{-0.77}$ \\
					$0.09\leq z<0.10$ &$37269$&  $10.56 ^{+0.32} _{-0.34}$ &  $10.64 ^{+0.32} _{-0.31}$ & $ -0.04 ^{+0.49} _{-1.27}$ &  $ -0.66^{+1.03} _{-0.75}$ & $-10.70^{+0.80} _{-1.41}$  & $-11.44 ^{+1.40} _{-0.81}$ \\
					$0.10\leq z<0.11$ &$37715$&  $10.63 ^{+0.29} _{-0.31}$ & $10.70^{+0.29} _{-0.29}$ &$ -0.08 ^{+0.58} _{-1.20}$&  $-0.71 ^{+1.13} _{-0.67}$& $-10.82 ^{+0.89} _{-1.32}$  & $-11.53 ^{+1.44} _{-0.74}$ \\
					$0.11\leq z<0.12$ &$38889$&  $10.72^{+0.27} _{-0.30}$ &  $10.78 ^{+0.26} _{-0.26}$& $ -0.17^{+0.70} _{-1.09}$ & $-0.69^{+1.15} _{-0.65}$& $-11.01^{+1.02} _{-1.17}$  & $-11.60^{+1.45} _{-0.69}$  \\
					$0.12\leq z<0.13$ &$33846$&  $10.79 ^{+0.24} _{-0.27}$& $10.85 ^{+0.23} _{-0.24}$ & $ -0.17 ^{+0.77} _{-1.06}$&  $-0.70^{+1.20} _{-0.60}$ & $ -11.07^{+1.07} _{-1.12}$  & $-11.67 ^{+1.47} _{-0.64}$ \\
					$0.13\leq z<0.14$ &$32447$& $10.86^{+0.22} _{-0.26}$ & $10.91^{+0.20} _{-0.22}$ & $ -0.27^{+0.89} _{-0.95}$ &  $-0.73 ^{+1.25} _{-0.55}$ & $ -11.25^{+1.18} _{-0.99}$  &$-11.75 ^{+1.47} _{-0.59}$  \\
					$0.14\leq z<0.15$ &$25878$&  $10.92 ^{+0.19} _{-0.24} $ & $10.97 ^{+0.19} _{-0.20}$   & $ -0.33 ^{+0.98} _{-0.87}$ &  $-0.76 ^{+1.28} _{-0.52}$ & $-11.36^{+1.24} _{-0.91}$  & $-11.82 ^{+1.48} _{-0.54}$ \\
					$0.15\leq z<0.16$ &$21074$& $10.98^{+0.17} _{-0.23}$ &  $11.02 ^{+0.16} _{-0.19}$ & $ -0.35 ^{+1.04} _{-0.84}$& $-0.74 ^{+1.29} _{-0.51}$  & $-11.44 ^{+1.28} _{-0.86}$  & $-11.86^{+1.47} _{-0.52}$ \\
					$0.16\leq z<0.17$ &$16899$& $11.04 ^{+0.15} _{-0.22}$ &  $11.08 ^{+0.14} _{-0.18}$ & $ -0.41^{+1.13} _{-0.76}$& $-0.73 ^{+1.32} _{-0.50}$  & $-11.54^{+1.35} _{-0.79}$  & $-11.89^{+1.46} _{-0.51}$ \\
					$0.17\leq z<0.18$ &$12967$& $11.09^{+0.14} _{-0.21}$ &  $11.12 ^{+0.12} _{-0.17}$ & $ -0.38 ^{+1.15} _{-0.76}$& $-0.69 ^{+1.30} _{-0.51}$  & $-11.56^{+1.36} _{ -0.77}$  & $-11.90 ^{+1.45} _{-0.52}$ \\
					$0.18\leq z<0.19$ &$9236$& $11.14^{+0.11} _{-0.21}$ &  $11.17^{+0.10} _{-0.16}$ & $ -0.41^{+1.22} _{-0.72}$& $-0.67 ^{+1.31} _{-0.51}$  & $ -11.63 ^{+1.42} _{-0.74}$  & $-11.92 ^{+1.44} _{-0.49}$ \\
					$0.19\leq z<0.2$ &$5734$& $11.18^{+0.11} _{-0.20}$ &  $11.21^{+0.09} _{-0.17}$ & $ -0.33 ^{+1.20} _{-0.76}$& $-0.63 ^{+1.31} _{-0.51}$  & $-11.61^{+1.42} _{-0.76}$  & $-11.89 ^{+1.44} _{-0.53}$ \\
					\bottomrule
				\end{tabular}
			}
			\label{mp}
	\end{table}
	\section{Conclusion}
	\label{secIV}
	In this study, for the first time we made the use of the 
	value-added 
	catalogue by \citet{tempel2017merging}, consisting of 584,449 
	galaxies 
	derived from SDSS DR12. The MPA-JHU measurements were used to investigate the impact of 
	isolated 
	and group environments on stellar mass, SFR, and SSFR. To mitigate 
	the Malmquist bias, 
	the entire sample was divided into eighteen sub-samples with 
	a redshift 
	binning size of $\Delta z = 0.01$. We analyzed the properties of galaxies
	sub-samples 
	in each redshift bin for both isolated and group environments within the 
	range of $0.02 \leq z < 0.2$. The results, as 
	depicted in Figures 
	\ref{stellarmass}, \ref{sfr},  \ref{ssfr}, \ref{zmasssfrssfr}, and detailed in 
	Table \ref{mp}, 
	provide the valuable insights. Additionally, we carried out the two sample KS, AD 
	tests between 
	isolated and group galaxies shown in Table \ref{st}, this results align 
	with the conclusions 
	drawn from Figures \ref{stellarmass}, \ref{sfr},  \ref{ssfr}, \ref{zmasssfrssfr}, 
	reinforcing 
	the statistical significance of the findings. The key findings of this study are:
\begin{table}[!h]
	\centering
    \caption{The KS, AD statistics and their corresponding p-values (in brackets) for the stellar mass (M$\star$), SFR and SSFR.}
		\vspace{8pt}
		\setlength{\tabcolsep}{0.5pc}
		\scalebox{1}{
			\begin{tabular}{cccc|ccc}
				\toprule
				\toprule
				& \multicolumn{3}{c}{KS statistics (P-values)} & \multicolumn{3}{c}{AD statistics (P-values)} \\
				\cmidrule(lr){2-4} \cmidrule(lr){5-7}
				Redshift & M$\star$ & SFR & SSFR & M$\star$ & SFR & SSFR\\
				{(1)} & {(2)} & {(3)} & {(4)} & {(5)} & {(6)} & {(7)} \\
				\midrule
				$0.02\leq z<0.03$ & 0.18 (1.04e-23) & 0.21 (1.24e-31) & 0.22 (6.02e-34) & 90.76 (1e-5) & 70.59 (1e-5) & 106.37 (1e-5) \\
				$0.03\leq z<0.04$ & 0.16 (3.80e-34) & 0.19 (1.83e-46) & 0.20 (1.52e-50) & 131.31 (1e-5) & 108.79 (1e-5) & 159.61 (1e-5) \\
				$0.04\leq z<0.05$ & 0.13 (1.83e-46) & 0.15 (1.34e-43) & 0.16 (1.34e-47) & 133.28 (1e-5) & 102.64 (1e-5) & 157.68 (1e-5) \\
				$0.05\leq z<0.06$ & 0.13 (6.67e-46) & 0.16 (4.95e-74) & 0.17 (3.17e-78) & 190.69 (1e-5) & 180.48 (1e-5) & 257.60 (1e-5)\\
				$0.06\leq z<0.07$ & 0.12 (9.83e-66) & 0.15 (1.57e-110) & 0.15 (8.63e-111) & 304.82 (1e-5) & 277.55 (1e-5) & 398.71 (1e-5) \\
				$0.07\leq z<0.08$ & 0.12 (2.41e-101) & 0.15 (9.34e-160) & 0.16 (3.15e-170) & 448.83 (1e-5) & 419.60 (1e-5) & 587.52 (1e-5) \\
				$0.08\leq z<0.09$ & 0.11 (9.85e-92) & 0.15 (6.63e-191) & 0.16 (6.39e-197) & 384.70 (1e-5) & 565.64 (1e-5) & 704.41 (1e-5)\\
				$0.09\leq z<0.10$ & 0.40 (6.72e-89) & 0.14 (1.84e-154) & 0.14 (7.85e-165) & 394.52 (1e-5) & 436.67 (1e-5)& 568.64 (1e-5) \\
				$0.10\leq z<0.11$  & 0.10 (8.23e-85) & 0.14 (2.76e-156) & 0.14 (2.78e-165) & 375.15 (1e-5) & 472.25 (1e-5) & 595.56 (1e-5) \\
				$0.11\leq z<0.12$  & 0.10 (7.98e-84) & 0.13 (8.66e-126) & 0.13 (2.23e-138) & 337.11 (1e-5) & 379.83 (1e-5) & 484.76 (1e-5)\\
				$0.12\leq z<0.13$ & 0.11 (6.49e-83) & 0.13 (7.23e-118) & 0.13 (7.48e-127) & 358.43 (1e-5) & 369.28 (1e-5) & 466.96 (1e-5) \\
				$0.13\leq z<0.14$ & 0.11 (2.33e-86) & 0.12 (4.98e-100) & 0.13 (1.91e-108) & 342.62 (1e-5) & 323.31 (1e-5) & 413.38 (1e-5) \\
				$0.14\leq z<0.15$ & 0.10 (1.08e-57) & 0.13 (1.69e-84) & 0.13 (2.39e-90) & 223.09 (1e-5) & 286.47 (1e-5) & 334.25 (1e-5) \\
				$0.15\leq z<0.16$ & 0.10 (2.70e-43) & 0.13 (3.56e-66) & 0.13 (3.31e-68) & 169.85 (1e-5) & 207.24 (1e-5) & 240.55 (1e-5) \\
				$0.16\leq z<0.17$ & 0.11 (2.07e-38) & 0.12 (8.88e-48) & 0.12 (4.24e-48) & 136.31 (1e-5) & 136.02 (1e-5) & 159.02 (1e-5)\\
				$0.17\leq z<0.18$ & 0.11 (8.55e-28) & 0.12 (3.3e-29) & 0.12 (7.33e-31) & 91.30 (1e-5) & 96.82 (1e-5) & 112.04 (1e-5) \\
				$0.18\leq z<0.19$ & 0.11 (7.97e-16) & 0.12 (2.65e-17) & 0.12 (3.12e-18) & 54.57 (1e-5) & 58.84 (1e-5) & 64.23 (1e-5)\\
				$0.19\leq z<0.20$ & 0.10 (6.58e-12) & 0.11 (4.86e-14) & 0.11 (2.46e-16) & 36.29 (1e-5) & 30.38 (1e-5) & 33.95 (1e-5) \\
				\bottomrule
		\end{tabular}}
		\label{st}
\end{table}
	\begin{itemize}
		\item High-mass galaxies tend to exist in groups across all redshift 
		bins, while 
		low-mass galaxies show a preference for isolated environments. 
		This is 
		consistent with previous conclusions regarding galaxy distributions in 
		high and 
		low-density regions, that high-mass galaxies exist in the high-density 
		region of 
		the Universe while low-mass galaxies exist in the low-density region 
		of the Universe \cite{kauffmann2004environmental, li2006dependence}.
		\item Galaxies in group environments exhibit lower SFR and SSFR across 
		all redshift 
		bins, whereas galaxies in isolated environments tend to have 
		higher 
		SFR and SSFR. This aligns with the conclusion from other studies that 
		galaxies 
		display higher SFR, SSFR in low-density regions of the Universe, and 
		low SFR, 
		SSFR in higher-density regions of the Universe \cite{lewis2002df, 
			gomez2003galaxy, tanaka2004environmental, elbaz2007reversal, cooper2008deep2,patel2009dependence}.
		\item The proportion of galaxies within isolated and group environments
		depends on 
		the redshift values. In the lower redshift bins ($z \lesssim 0.1$), the 
		galaxy proportion 
		is higher in isolated than in the group 
		environments. 
		In the intermediate redshift bins ($0.1\lesssim z \lesssim 0.12$) galaxies 
		have almost 
		equal proportionality in both environments. Whereas in the higher
		redshift 
		bins ($z\gtrsim 0.12$) the proportionality of galaxies is higher in 
		the group 
		environments than in the isolated environments.
		\item The strong dependence of stellar mass, SFR and SSFR on environment was observed, where by this dependence is much stronger at lower redshift and decrease at higher redshift.
	\end{itemize}
	These findings contribute to a more profound comprehension of the intricate 
	interplay 
	between stellar mass, SFR, and SSFR, and galaxy environment using 
	clustering 
	approach through comparing isolated and group environment. The 
	study not 
	only reveals patterns in the distribution of these properties across different 
	redshift 
	bins but also underscores the significant impact that clustering have 
	on the observed 
	characteristics of galaxies in comparison to the existing studies 
	which were done using galaxy density approach.
	
	In a future paper we will examine the evolution of the physical properties 
	(e.g., stellar mass and star formation rate) for emission line and morphological 
	classified galaxies in different local environments (e.g., isolated, group and cluster), 
	and interpret the results more quantitatively studying the influence of AGN on the observed properties.
	
	\section*{Acknowledgements} PP acknowledges support from The Government of Tanzania 
	through the India Embassy, Mbeya University of Science and Technology (MUST) 
	for Funding and SDSS for providing data. UDG is thankful to the Inter-University 
	Centre for Astronomy and Astrophysics (IUCAA), Pune, India for the Visiting 
	Associateship of the institute.


\begin{thebibliography}{99}
		
		\bibitem[Terrazas et al.(2016)]{terrazas2016quiescence}
		B. Terrazas, E. Bell, B. Henriques, S. White, A. Cattaneo, J. Woo, \emph{Quiescence correlates strongly with directly-measured black hole mass in central galaxies}, \href{https://doi.org/10.3847/2041-8205/830/1/L12}{The ApJL \textbf{830}, L12 (2016)} [\href{https://arxiv.org/abs/1609.07141}{arXiv:1609.07141}].
		
		\bibitem[Peng et al.(2010)] {peng2010mass}
		Y. Peng, S. Lilly, K. Kovač, M. Bolzonella, L. Pozzetti, A. Renzini, G. Zamorani,  \emph{Mass and Environment as Drivers of Galaxy Evolution in SDSS and zCOSMOS and the Origin of the Schechter Function}, \href{https://doi.org/10.1088/0004-637X/721/1/193}{ApJ \textbf{721}, 193 (2010)} [\href{https://arxiv.org/abs/1003.4747}{arXiv:1003.4747}].
		
		\bibitem[Cooper et al.(2006)]{cooper2006deep2}
		M. Cooper, J. Newman,  \emph{The DEEP2 Galaxy Redshift Survey: The Relationship Between Galaxy Properties and Environment at $z \sim 1$}, \href{https://doi.org/10.1111/j.1365-2966.2006.10485.x}{MNRAS \textbf{370}, 198 (2006)} [\href{https://arxiv.org/abs/astro-ph/0603177}{arXiv:astro-ph/0603177}].	
		
		\bibitem[Yoon et al.(2023)]{yoon2023low}
		Y. Yoon, J. Kim,  \emph{Low-mass Quiescent Galaxies Are Small in Isolated Environments: Environmental Dependence of the Mass-Size Relation of Low-mass Quiescent Galaxies}, \href{https://doi.org/ 10.3847/1538-4357/acfed5}{ApJ\textbf{957}, 59 (2023)}   [\href{https://arxiv.org/abs/2310.07498}{arXiv:2310.07498}].
		
		\bibitem[Sankhyayan et al.(2023)]{sankhyayan2023identification}
		S. Sankhyayan, J. Bagchi,\emph{Identification of Superclusters and their Properties in the Sloan Digital Sky Survey Using WHL Cluster Catalog}, \href{https://doi.org/10.3847/1538-4357/acfaeb}{ApJ \textbf{958}, 62 (2023)}   [\href{https://arxiv.org/abs/2309.06251}{arXiv:2309.06251}].
		
		\bibitem[Bag et al.(2023)]{bag2023shape}
		S. Bag, L. Liivamägi, M. Einasto,\emph{The shape distribution of superclusters in SDSS DR 12}, \href{https://doi.org/10.1093/mnras/stad811}{MNRAS \textbf{521},  4712 (2023)}   [\href{https://arxiv.org/abs/2111.10253}{arXiv:2111.10253}].
		
		\bibitem[Grutzbauch et al.(2011)]{grutzbauch2011galaxy}
		R. Gr\"utzbauch, R. Chuter, \emph{Galaxy properties in different environments up to $z \sim 3$ in the GOODS NICMOS Survey}, \href{https://doi.org/10.1111/j.1365-2966.2010.18060.x}{MNRAS \textbf{412}, 2361 (2011)} [\href{https://arxiv.org/abs/1011.4846}{arXiv:1011.4846}].	
		
		\bibitem[Ball et al.(2008)]{ball2008galaxy}
		N. Ball, J. Loveday, \emph{Galaxy Colour, Morphology, and Environment in the Sloan Digital Sky Survey}, \href{https://doi.org/10.1111/j.1365-2966.2007.12627.x}{MNRAS \textbf{383},  907 (2008)}   [\href{https://arxiv.org/abs/astro-ph/0610171}{arXiv:astro-ph/0610171}].
		
		\bibitem[Kauffman et al.(2004)]{kauffmann2004environmental}
		G. Kauffmann, S. White, \emph{The Environmental Dependence of the Relations between Stellar Mass, Structure, Star Formation and Nuclear Activity in Galaxies}, \href{https://doi.org/10.1111/j.1365-2966.2004.08117.x}{MNRAS \textbf{353}, 713 (2004)} [\href{https://arxiv.org/abs/astro-ph/0402030}{arXiv:astro-ph/0402030}].
		
		\bibitem[Etherington et al.(2017)]{etherington2017environmental}
		J. Etherington, D. Thomas, \emph{Environmental dependence of the galaxy stellar mass function in the Dark Energy Survey Science Verification Data}, \href{https://doi.org/10.1093/mnras/stw3069}{A\& A \textbf{466}, 228 (2016)} [\href{https://arxiv.org/abs/1701.06066}{arXiv:1701.06066}].
		
		\bibitem[Li et al.(2006)]{li2006dependence}
		C. Li, G. Kauffmann, Y. Jing, S. White, \emph{The dependence of clustering on galaxy properties}, \href{https://doi.org/10.1111/j.1365-2966.2006.10066.x}{MNRAS \textbf{368}, 21 (2006)} [\href{https://arxiv.org/abs/astro-ph/0509873}{arXiv:astro-ph/0509873}].
		
		\bibitem[Scudder et al.(2012)]{scudder2012dependence}
		J. Scudder, S. Ellison,\emph{The dependence of galaxy group star formation rates and metallicities on large-scale environment}, \href{https://doi.org/10.1111/j.1365-2966.2012.21080.x}{ApJ \textbf{423}, 2690 (2012)}   [\href{https://arxiv.org/abs/1204.2828}{arXiv:1204.2828}].
		
		\bibitem[Gomez et al.(2003)]{gomez2003galaxy}
		P. Gomez, R. Nichol, \emph{Galaxy Star-Formation as a Function of Environment in the Early Data Release of the Sloan Digital Sky Survey}, \href{https://doi.org/10.1086/345593}{A \& A \textbf{584}, 210 (2003)} [\href{https://arxiv.org/abs/astro-ph/0210193}{arXiv:astro-ph/0210193}].
		
		\bibitem[Lewis et al.(2002)]{lewis2002df}
		I. Lewis, M. Balogh, R. Propris, W. Couch, R. Bower,  \emph{The 2dF Galaxy Redshift Survey: the environmental dependence of galaxy star formation rates near clusters}, \href{https://doi.org/10.1046/j.1365-8711.2002.05558.x}{MNRAS \textbf{334}, 673 (2002)} [\href{https://arxiv.org/abs/astro-ph/0203336}{arXiv:astro-ph/0203336}].
		
		\bibitem[Patel et al.(2009)]{patel2009dependence}
		S. Patel, B. Holden, \emph{The Dependence of Star Formation Rates on Stellar Mass and Environment at $z\sim 0.8$}, \href{https://doi.org/10.48550/arXiv.0910.0837}{ApJ \textbf{705}, L67 (2007)}   [\href{https://arxiv.org/abs/0910.0837}{arXiv:0910.0837}].
		
		\bibitem[Tanaka et al.(2004)]{tanaka2004environmental}
		M. Tanaka, T. Goto, \emph{The Environmental Dependence of Galaxy Properties in the Local Universe: Dependence on Luminosity, Local Density, and System Richness}, \href{https://doi.org/10.1086/425529}{AJ \textbf{128},  2677 (2004)}   [\href{https://arxiv.org/abs/astro-ph/0411132}{arXiv:astro-ph/0411132}].
		
		\bibitem[Cooper et al.(2008)]{cooper2008deep2}
		M. Cooper, J. Newman,  \emph{The DEEP2 Galaxy Redshift Survey: The Role of Galaxy Environment in the Cosmic Star-Formation History}, \href{https://doi.org/10.1111/j.1365-2966.2007.12613.x}{MNRAS \textbf{383}, 1058 (2008)} [\href{https://arxiv.org/abs/0706.4089}{arXiv:0706.4089}].
		
		\bibitem[Elbaz et al.(2007)]{elbaz2007reversal}
		D. Elbaz, E. Daddi, D. Borgne, \emph{The reversal of the star formation-density relation in the distant universe}, \href{https://doi.org/10.1051/0004-6361:20077525}{A\& A \textbf{468}, 33 (2007)}   [\href{https://arxiv.org/abs/astro-ph/0703653}{arXiv:astro-ph/0703653}].
		
		\bibitem[Deng et al.(2021)]{xin2021environmental}
		X. Deng, Z. Wu, \emph{Environmental Dependence of All the Five Band Luminosities of Active Galactic Nucleus (AGN) Host Galaxies}, \href{https://doi.org/10.1007/s10511-021-09706-y}{Astrophysics \textbf{64}, 446 (2021)}
		
		\bibitem[Deng et al.(2023)]{deng2023environmental}
		X. Deng, Z. Wu, \emph{Environmental Dependence of Different Colors of Active Galactic Nucleus (AGN) Host Galaxies}, \href{https://doi.org/10.1007/s10511-023-09790-2}{Astrophysics \textbf{66}, 173 (2023)}
		
		\bibitem[Rasmussen et al.(2012)]{rasmussen2012suppression}
		J. Rasmussen, J. S. Mulchaey, L. Bai, T. J. Ponman et al., \emph{The Suppression of Star Formation and the Effect of the Galaxy Environment in Low-redshift Galaxy Groups}, \href{ https://doi.org/10.1088/0004-637X/757/2/122}{ApJ \textbf{757}, 122 (2012)}
		
		\bibitem[Ziparo et al.(2013)]{ziparo2013lack}
		F. Ziparo, P. Popesso, A. Biviano, A. Finoguenov et al., \emph{The lack of star formation gradients in galaxy groups up to $z \sim 1.6$}, \href{ https://doi.org/10.1093/mnras/stt1222}{MNRAS \textbf{434}, 3089 (2013)}
		
		\bibitem[Wijesinghe et al.(2012)]{wijesinghe2012galaxy}
		D. B. Wijesinghe, A. M. Hopkins, S. Brough, E. N. Taylor et al., \emph{Galaxy And Mass Assembly (GAMA): galaxy environments and star formation rate variations}, \href{ https://doi.org/10.1111/j.1365-2966.2012.21164.x}{MNRAS \textbf{423}, 3679 (2012)}
		
		\bibitem[Schaefer et al.(2017)]{schaefer2017sami}
		A. L. Schaefer, S. M. Croom, J. T. Allen, S. Brough et al., \emph{The SAMI Galaxy Survey: spatially resolving the environmental quenching of star formation in GAMA galaxies}, \href{ https://doi.org/10.1093/mnras/stw2289}{MNRAS \textbf{464}, 121 (2017)}
		
		\bibitem[Wetzel et al.(2013)]{wetzel2013galaxy}
		A. R. Wetzel, J. L. Tinker, C. Conroy, F. C. Van Den Bosch et al., \emph{Galaxy evolution in groups and clusters: satellite star formation histories and quenching time-scales in a hierarchical Universe}, \href{ https://doi.org/10.1093/mnras/stt469}{MNRAS \textbf{432}, 336 (2013)}
		
		\bibitem[Wetzel et al.(2014)]{wetzel2014galaxy}
		A. R. Wetzel, J. L. Tinker, C. Conroy, F. C. Bosch et al., \emph{Galaxy evolution near groups and clusters: ejected satellites and the spatial extent of environmental quenching}, \href{ https://doi.org/10.1093/mnras/stu122}{MNRAS \textbf{439}, 2687 (2014)}
		
		\bibitem[McGee et al.(2014)]{mcgee2014overconsumption}
		S. L. McGee, R. G. Bower, M. L. Balogh, \emph{Overconsumption, outflows and the quenching of satellite galaxies}, \href{ https://doi.org/10.1093/mnrasl/slu066}{MNRAS \textbf{442}, L105 (2014)}
		
		\bibitem[Peng et al.(2015)]{peng2015strangulation}
		Y. Peng, R. Maiolino, R. Cochrane et al., \emph{Strangulation as the primary mechanism for shutting down star formation in galaxies}, \href{ https://doi.org/10.1093/mnrasl/slu066}{Nature \textbf{521}, 192 (2015)}
		
		\bibitem[McGee et al.(2014)]{grootes2017galaxy}
		M. W. Grootes, R. J. Tuffs, C. C. Popescu, P. Norberg et al., \emph{Galaxy And Mass Assembly (GAMA): gas fueling of spiral galaxies in the local universe. I. The effect of the group environment on star formation in spiral galaxies}, \href{ https://doi.org/10.3847/1538-3881/153/3/111}{AJ \textbf{153}, 111 (2017)}
		
		\bibitem[Guglielmo et al.(2015)]{guglielmo2015star}
		V. Guglielmo, B. M. Poggianti, A. Moretti, J. Fritz et al., \emph{The star formation history of galaxies: the role of galaxy mass, morphology and environment}, \href{ https://doi.org/10.1093/mnras/stv757}{MNRAS \textbf{450}, 2749 (2015)}
		
		\bibitem[Vulcani et al.(2014)]{vulcani2014bluer}
		B. Vulcani, B. M. Poggianti, J. Fritz, G. Fasano,  et al., \emph{From blue star-forming to red passive: galaxies in transition in different environmentst}, \href{ https://doi.org/10.1088/0004-637X/798/1/52}{ApJ \textbf{798}, 52 (2014)}
		
		\bibitem[Oemler et al.(2014)]{oemler2017star}
		A. Oemler, L. E. Abramson, M. D. Gladders, A. Dressler et al., \emph{The Star Formation Histories of Disk Galaxies: The Live, the Dead, and the Undead}, \href{ https://doi.org/10.3847/1538-4357/aa789e}{ApJ \textbf{844}, 45 (2017)}
		
		\bibitem[Barsanti et al.(2018)]{barsanti2018galaxy}
		S. Barsanti, M. S. Owers, S. Brough, L. J. M. Davies et al., \emph{Galaxy and mass assembly (GAMA): impact of the group environment on galaxy star formation}, \href{ https://doi.org/10.3847/1538-4357/aab61ae}{ApJ \textbf{857}, 71 (2018)}
		
		\bibitem[O'Neil et al.(2023)]{o2023searching}
		K. O'Neil, S. E. Schneider, W. van Driel et al., \emph{Searching in H i for Massive Low Surface Brightness Galaxies: Samples from HyperLeda and the UGC}, \href{ https://doi.org/10.3847/1538-3881/acd345}{AJ \textbf{165}, 263 (2023)}
		
		\bibitem[Tempel et al.(2017)]{tempel2017merging}
		E. Tempel, T. Tuvikene, \emph{Merging groups and clusters of galaxies from the SDSS data. The catalogue of groups and potentially merging systems}, \href{https://doi.org/10.1051/0004-6361/201730499}{A\& A \textbf{602}, A100 (2017)} [\href{https://arxiv.org/abs/1704.04477}{arXiv:1704.04477}]. 
		
		\bibitem[Ade et al.(2016)]{collaborartion2016planck}
		P. Ade, N. Aghanim,  \emph{Planck 2015 results. XXIII. The thermal Sunyaev-Zeldovich effect--cosmic infrared background correlation}, \href{https://doi.org/10.1051/0004-6361/201527418}{A\& A \textbf{594}, A23 (2016)} [\href{https://arxiv.org/abs/1509.06555}{arXiv:1509.06555}].
		
		\bibitem[Eisenstein et al.(2011)]{eisenstein2011sdss}
		D. Eisenstein, D. Weinberg, \emph{SDSS-III: Massive spectroscopic surveys of the distant universe, the Milky Way, and extra-solar planetary systems}, \href{https://doi.org/10.1088/0004-6256/142/3/72}{AJ \textbf{142}, 72 (2011)} [\href{https://arxiv.org/abs/1101.1529}{arXiv:1101.1529}].
		
		\bibitem[Alam et al.(2015)]{alam2015eleventh}
		S. Alam, F. Albareti, C. Prieto, \emph{The Eleventh and Twelfth Data Releases of the Sloan Digital Sky Survey: Final Data from SDSS-III}, \href{https://doi.org/10.1088/0067-0049/219/1/12}{ApJSS \textbf{219}, 12 (2015)} [\href{https://arxiv.org/abs/1501.00963}{arXiv:1501.00963}].
		
		\bibitem[Accetta et al.(2022)]{accetta2022seventeenth}
		K. Accetta, N. Abdurro'uf, C. Aerts, \emph{The Seventeenth Data Release of the Sloan Digital Sky Surveys: Complete Release of MaNGA, MaStar and APOGEE-2 Data}, \href{https://doi.org/10.3847/1538-4365/ac4414}{ApJSS \textbf{259}, 1 (2022)} [\href{https://arxiv.org/abs/2112.02026}{arXiv:2112.02026}].
		
		\bibitem[Tempel et al.(2014)]{tempel2014flux}
		E. Tempel, A. Tamm, M. Gramann, \emph{Flux- and volume-limited groups/clusters for the SDSS galaxies: catalogues and mass estimation}, \href{https://doi.org/10.1051/0004-6361/201423585}{A\& A \textbf{566}, A1 (2014)}   [\href{https://arxiv.org/abs/1402.1350}{arXiv:1402.1350}].
		
		\bibitem[Blanton et al.(2007)]{blanton2007k}
		M. Blanton, S. Roweis, \emph{K-corrections and filter transformations in the ultraviolet, optical, and near-infrared}, \href{https://doi.org/10.1086/510127}{AJ \textbf{133}, 734 (2007)} [\href{https://arxiv.org/abs/astro-ph/0606170}{arXiv:astro-ph/0606170}].
		
		\bibitem[Blanton et al.(2003)]{blanton2003galaxy}
		M. Blanton, D. Hogg, N. Bahcall, J. Brinkmann et al., \emph{The galaxy luminosity function and luminosity density at redshift $z= 0.1$}, \href{https://doi.org/ 10.1086/375776}{AJ\textbf{592}, 819 (2003)} [\href{https://arxiv.org/abs/astro-ph/0210215}{arXiv:astro-ph/0210215}].
		
		\bibitem[Tempel et al.(2012)]{tempel2012groups}
		E. Tempel, E. Tago, L. Liivam\"agi, \emph{Groups and clusters of galaxies in the SDSS DR8-Value-added catalogues}, \href{https://doi.org/10.1051/0004-6361/201118687}{A \& A \textbf{540}, A106 (2012)}   [\href{https://arxiv.org/abs/1112.4648}{arXiv:1112.4648}].
		
		\bibitem[Brinchmann et al.(2004)]{brinchmann2004physical}
		J. Brinchmann, S. Charlot, \emph{The physical properties of star forming galaxies in the low redshift universe}, \href{https://doi.org/10.1111/j.1365-2966.2004.07881.x}{MNRAS \textbf{351}, 1151 (2004)} [\href{https://arxiv.org/abs/astro-ph/0311060}{arXiv:astro-ph/0311060}].
		
		\bibitem[Tremonti et al.(2004)]{tremonti2004origin}
		C. Tremonti, T. Heckman,\emph{The Origin of the Mass--Metallicity Relation: Insights from 53,000 Star-Forming Galaxies in the SDSS}, \href{https://doi.org/10.1086/423264}{ApJ \textbf{613},  898 (2004)}   [\href{https://arxiv.org/abs/astro-ph/0405537}{arXiv:astro-ph/0405537}].
		
		\bibitem[York et al.(2000)]{york2000sloan}
		D. York, J. Adelman, \emph{The sloan digital sky survey: Technical summary}, \href{https://doi.org/10.1086/301513}{AJ \textbf{120},  1579 (2000)}   [\href{https://arxiv.org/abs/astro-ph/0006396}{arXiv:astro-ph/0006396}]
		
		\bibitem[Salim et al.(2015)]{salim2015mass}
		S. Salim, J. Lee, R. Davé,  \emph{On the Mass-Metallicity-Star Formation Rate Relation for Galaxies at $ z\sim 2$}, \href{https://doi.org/10.1088/0004-637X/808/1/25}{ApJ \textbf{808}, 14pp (2015)} [\href{https://arxiv.org/abs/1506.03080}{arXiv:1506.03080}].
		
		\bibitem[Teerikorp  et al.(2015)]{teerikorpi2015eddington}
		P. Teerikorp, \emph{Eddington-Malmquist bias in a cosmological context}, \href{https://doi.org/10.1051/0004-6361/201425489}{A\& A \textbf{576},  A75 (2015)}   [\href{https://arxiv.org/abs/1503.02812}{arXiv:1503.02812}].
		
		\bibitem[Hodges.(1958)]{hodges1958significance}
		J. L. Hodges, \emph{The significance probability of the Smirnov two-sample test}, \href{https://doi.org/ 10.1007/BF02589501}{Arkiv f{\"o}r matematik \textbf{3},  469 (1958)}   
		
		\bibitem[Harari and Mollerach.(2009)]{harari2009kolmogorov}
		D. Harari, S. Mollerach, \emph{Kolmogorov-Smirnov test as a tool to study the distribution of ultra-high energy cosmic ray sources}, \href{https://doi.org/10.1111/j.1365-2966.2008.14327.x}{A \& A \textbf{394},  916 (2009)}   [\href{https://arxiv.org/abs/0811.0008}{arXiv:0811.0008}].
		
		\bibitem[Babu and Feigelson.(2006)]{babu2006astrostatistics}	
		G. J. Babu, E. D. Feigelson, \emph{Astrostatistics: Goodness-of-fit and all that},  \href{2006ASPC..351..127B} {ADASS XV \textbf{351},  127 (2006)}.	
	\end{thebibliography}
\end{document}